%% file: wzw_Journal.tex
\documentclass[final]{IEEEtran}

\usepackage{newclude} 
\usepackage{cite}     
\usepackage{color}
\usepackage{threeparttable}
\usepackage{graphicx}
\usepackage{picinpar}
\usepackage{epstopdf}
\usepackage[cmex10]{amsmath}
\usepackage{amsmath,amsfonts,amssymb}
\usepackage{subfigure}
\usepackage{changepage}
\usepackage{algorithm}
\usepackage{caption}
\usepackage{algpseudocode}
\usepackage{stfloats}
\usepackage{bm}
\usepackage{amsthm}
\usepackage{enumerate}
\usepackage{multirow}
\usepackage{booktabs}
\usepackage{url}
\usepackage{balance}

\usepackage[colorlinks=true, citecolor=blue, linkcolor=blue, urlcolor=blue]{hyperref}
\title{Ultra-Massive MIMO with Orthogonal Chirp Division Multiplexing for Near-Field Sensing and Communication Integration}

\author{
Ziwei~Wan,~\IEEEmembership{Graduate Student Member,~IEEE},
Zhen~Gao,~\IEEEmembership{Senior Member,~IEEE},
Fabien~H\'eliot,~\IEEEmembership{Senior Member,~IEEE},
Qu~Luo,~\IEEEmembership{Member,~IEEE},
Pei~Xiao,~\IEEEmembership{Senior Member,~IEEE},
Haiyang~Zhang,~\IEEEmembership{Member,~IEEE},
Christos~Masouros,~\IEEEmembership{Fellow,~IEEE},
Yonina C. Eldar,~\IEEEmembership{Fellow,~IEEE},
and Sheng~Chen,~\IEEEmembership{Life Fellow,~IEEE}

}

\begin{document}

\newtheorem{remark}{Remark}
\newtheorem{lemma}{Lemma}
\newtheorem{corollary}{Corollary}
\newtheorem{theorem}{Theorem}
\newtheorem{proposition}{Proposition}
\newtheorem{definition}{Definition}
\newtheorem{condition}{Condition}

\definecolor{red}{rgb}{0,0,0}

\maketitle

\include*{Sec/Abstract}

\include*{Sec/Sec1_Intro}
\include*{Sec/Sec2_Preliminaries}
\include*{Sec/Sec3_Model}
\include*{Sec/Sec4_Sensing}

\include*{Sec/Sec5_Communications}
\include*{Sec/Sec6_Simulation}
\include*{Sec/Sec7_Conclusion}

\include*{Sec/Appendix}

\bibliographystyle{IEEEtran}
\bibliography{References}

\balance

\end{document}

%% file: Sec/Abstract.tex
\begin{abstract}
This paper integrates the emerging ultra-massive multiple-input multiple-output (UM-MIMO) technique with orthogonal chirp division multiplexing (OCDM) waveform to tackle the challenging near-field integrated sensing and communication (ISAC) problem. Specifically, we conceive a comprehensive ISAC architecture, where an UM-MIMO base station adopts OCDM waveform for communications and a co-located sensing receiver adopts the frequency-modulated continuous wave (FMCW) detection principle to simplify the associated hardware. For sensing tasks, several OCDM subcarriers, namely, dedicated sensing subcarriers (DSSs), are each transmitted through a dedicated sensing antenna (DSA) within the transmit antenna array. By judiciously designing the DSS selection scheme and optimizing receiver parameters, the FMCW-based sensing receiver can decouple the echo signals from different DSAs with significantly reduced hardware complexity. This setup enables the estimation of ranges and velocities of near-field targets in an antenna-pairwise manner. Moreover, by leveraging the spatial diversity of UM-MIMO, we introduce the concept of virtual bistatic sensing (VIBS), which incorporates the estimates from multiple antenna pairs to achieve high-accuracy target positioning and three-dimensional velocity measurement. The VIBS paradigm is immune to hostile channel environments characterized by spatial non-stationarity and uncorrelated multipath environment. Furthermore, the channel estimation of UM-MIMO OCDM systems enhanced by the sensing results is investigated. Simulation results demonstrate that the proposed ISAC scheme enhances sensing accuracy, and also benefits communication performance.
\end{abstract}


%% file: Sec/Sec1_Intro.tex
\section{Introduction}\label{sec:1}

Recently, the International Telecommunication Union Radiocommunication Sector (ITU-R) has outlined a roadmap for the sixth-generation (6G) research \cite{liu2023vision,wan2024orthogonal},
where integrated sensing and communication (ISAC) is recognized as a crucial component of 6G wireless to unlock transformative future opportunities \cite{liu2020joint}.
Current research on ISAC has generally focused on two key aspects: the {\em spatial domain} and the {\em time-frequency domain}. The spatial domain research emphasizes the design of antenna arrays for ISAC, which enhances spatial degrees of freedom to boost both communication rates and sensing accuracy \cite{gao2022integrated}. The time-frequency domain research, on the other hand, concentrates on optimizing waveforms in ISAC systems to fully utilize the limited time and bandwidth resources \cite{zhou2022integrated}. While significant progress has been achieved in each of these areas independently, exploring their interaction is anticipated to drive exciting advancements of future ISAC.

In this paper, we introduce an innovative ISAC approach that involves both spatial domain and time-frequency domain.
The emerging ultra-massive multiple-input multiple-output (UM-MIMO) and orthogonal chirp division multiplexing (OCDM) techniques are incorporated to facilitate near-field ISAC, which has not been reported in previous literature. 

\subsection{Related Works}\label{S1.1}

We review the literature regarding the ISAC design in the spatial domain and time-frequency domain respectively.

\subsubsection{Spatial domain}

Massive multiple-input multiple-output (mMIMO) plays a pivotal role in ISAC \cite{gao2022integrated,peng2025latency}.
However, previous studies on mMIMO-aided ISAC predominantly rely on the {far-field} channel model.
To fulfill the more stringent performance requirements in 6G, the antenna array is evolving from mMIMO to UM-MIMO \cite{xiao2023sm,xiao2023error}, which incorporates an extensive number of antennas to form a very large aperture, thereby inducing the {near-field} effect \cite{cui2022near}.
A key characteristic of near-field effect is the {spherical wavefront} of electromagnetic wave, which introduces an additional distance dimension in the spatial domain for both sensing and communications \cite{cong2024near,wang2023near}.
This shift has made near-field ISAC a timely and increasingly popular research area.
For instance, the authors of \cite{wang2023near} proposed a near-field ISAC framework aided by UM-MIMO, achieving globally optimal solutions for both fully digital arrays and hybrid arrays by minimizing the Cram\'er-Rao bound (CRB). A similar scenario was explored in \cite{galappaththige2024near}, where the transmit power of ISAC station was minimized while fulfilling communication and sensing rate requirements. 
Furthermore, the authors of \cite{wang2024cramer} conducted the CRB analysis for near-field sensing with UM-MIMO, which validates the superiority of the near-field ISAC over far-field approaches. However, these works \cite{wang2023near,galappaththige2024near,wang2024cramer} primarily focused on narrowband scenarios, while the wideband near-field ISAC is deemed to be more powerful to support various applications \cite{wang2024rethinking}.
Research on wideband near-field ISAC is still at its infancy. 
The authors of \cite{wang2024wideband} studied the wideband near-field ISAC systems, in which the precoding matrix and the antenna selection strategies were optimized to balance the performance tradeoff between sensing and communication. In \cite{wang2024performance}, the narrowband results from \cite{wang2024cramer} was extended to wideband scenarios, where the CRB of the wideband near-field ISAC was derived and analyzed.

{\color{red}
In addition, to fulfill the more stringent performance requirements in 6G, the spatial-domain design is evolving towards antenna position flexibility , which is represented by emerging architectures like movable antenna (MA) \cite{zhu2026tut,sun2026movable}, fluid antenna system (FAS) \cite{wong2021fluid}, and the six-dimensional movable antenna (6DMA) \cite{shao20256d,shao2025JSAC}. By enabling the quasi-continuous movement or rotation of antenna elements, these technologies can proactively reconfigure the wireless channel to optimize signal alignment and interference mitigation. 
}

\subsubsection{Time-frequency domain (Waveform)}

Waveform design is a crucial aspect of both communications and sensing, and integrated waveform design represents a milestone technology for realizing ISAC \cite{zhou2022integrated,feng2020joint}.
The most representative waveform is orthogonal frequency division multiplexing (OFDM), which is 
widely considered in ISAC \cite{sturm2011waveform,liu2017adaptive,huang2022designing,wang2024performance}.
Another waveform candidate for ISAC is orthogonal time frequency space (OTFS) \cite{hadani2017orthogonal}, which employs the delay-Doppler domain to transmit the data.
Studies have been conducted on OTFS-based communications \cite{hadani2017orthogonal,wei2021orthogonal}, radar sensing \cite{raviteja2019orthogonal}, and ISAC \cite{gaudio2020effectiveness,wu2021otfs,yuan2021integrated}.
More recently, the OCDM has emerged as a promising waveform for wireless applications.
The OCDM was proposed in \cite{ouyang2016orthogonal}, where its effectiveness in wireless communications was initially verified. The authors of \cite{omar2021performance} studied the communication performance of OCDM under various channel conditions and compared it with the traditional OFDM and single carrier schemes.
In \cite{zhang2021channel}, an OCDM channel estimation (CE) method was introduced to address carrier frequency offset.
The CE for OCDM was further studied under MIMO scenario in \cite{ouyang2023channel} by leveraging the convolution-preserving property of the Fresnel transforms.
Due to the strong resistance against Doppler effect, OCDM has also been widely adopted in underwater acoustic communications \cite{wang2023underwater,jia2024ocdm}.
In the context of ISAC, OCDM constitutes a compelling solution by enabling joint communications and sensing through the shared chirp waveforms.
The authors of \cite{bhattacharjee2022evaluation} explored a bistatic vehicular ISAC system that leverages OCDM, achieving both high range resolution and a high communication rate through the sequential symbol decoding and radar parameter estimation algorithm. A comparative study \cite{de2021joint} examined the ISAC performance of OCDM against other state-of-the-art schemes. An OCDM-ISAC system operating in the terahertz band was introduced in \cite{li2023thz}, to highlight its potential for ultra-high frequency applications. In \cite{li2024orthogonal}, index modulation was integrated into OCDM-aided ISAC systems, to enable the information transmission while maintaining sensing performance.

\subsection{Motivation and Contributions}\label{S1.2}

Based on the aforementioned reviews, it is evident that a holistic ISAC design encompassing both the spatial and time-frequency domains remains underexplored. 
In addition, most of ISAC research overlooks the extra complexity imposed by the sensing receivers.
For example, the scheme proposed in \cite{wang2024performance} requires a sensing receiver equipped with $512$ power-hungry radio frequency (RF) chains, each needing an analog-to-digital converter (ADC) with sampling rate not lower than the system bandwidth. This poses challenges in practical deployment.
To tackle these issues, we improve the near-field ISAC systems by offering higher sensing accuracy, stronger robustness to diversified channels, and lower hardware complexity.
Our main contributions can be summarized as follows. 

\begin{itemize}
    \item {\bf We conceive an UM-MIMO-aided ISAC architecture combining the OCDM and frequency-modulated continuous wave (FMCW) detection techniques.} The proposed architecture consists of an OCDM transmitter with UM-MIMO and a sensing receiver with FMCW filterbanks. At the sensing stage, we transmit a few OCDM subcarriers, namely, dedicated sensing subcarriers (DSSs), each through a dedicated sensing antenna (DSA) of the UM-MIMO. Based on the proposed DSS selection scheme and receiver parameter design, the sensing receiver can effectively decouple the echo signals from different transmit-receive antenna pairs with significantly reduced hardware complexity, laying the foundation for subsequent processing.

    \item {\bf We propose a parameter estimation method taking the dual discontinuity into account.} It is revealed that the intermediate frequency (IF) signals at the sensing receiver may exhibit dual discontinuity when the spectrum-folded OCDM subcarriers are processed by the FMCW receiver, which is rarely considered in previous studies and invalidates many existing parameter estimation methods. As a remedy, we combine the reduced-length ESPRIT (RL-ESPRIT) and gradient descent algorithm (GDA) to realize the super-resolution estimation of the ranges and velocities for each antenna pair under dual discontinuity. 
    
    \item {\bf We introduce the concept of virtual bistatic sensing (VIBS) for the near-field target positioning and velocity measurements.} When UM-MIMO is deployed, the range and velocity of the near-field target observed by different antennas will be significantly distinct, even though all antennas are actually located at a single station. This spatial diversity motivates us to incorporate the measurements from different antenna pairs to determine the position and three-dimensional (rather than only radial) velocity of the target, which is called VIBS. We show that compared to the existing schemes, VIBS is more robust to the hostile channel environments.

    \item {\bf A sensing-enhanced near-field communication CE scheme for UM-MIMO with OCDM is proposed.} The near-field communication CE is formulated as the compressive sensing (CS) problem under OCDM waveform and the initial result is obtained via distributed orthogonal matching pursuit (DOMP). Then, the sensing results are introduced to refine the support of near-field channels in the polar-domain to enhance CE. This validates the concept of ISAC by benefiting the communication tasks via sensing. 
\end{itemize}

\subsection{Notations}\label{S1.3}

Column vectors and matrices are denoted by lower- and upper-case boldface letters, respectively. $(\cdot )^{*}$, $(\cdot )^{\rm T}$, $(\cdot )^{\rm H}$, and $(\cdot )^{\dag}$ denote the conjugate, transpose, conjugate transpose, and pseudo-inverse  operators, respectively. 
$\left\lfloor \cdot \right\rfloor$ represents the ceiling operator.
${\bf I}_N$ is the $N\times N$ identity matrix. 
The cardinality of a set $\cal I$ is denoted by $\left| {\cal I} \right|_{\rm c}$.
$[{\bf a}]_{i}$ is the $i$-th element of $\bf{a}$, $\left[{\bf A}\right]_{i,j}$ is the $i$-th row and $j$-th column element of matrix $\bf{A}$, ${\rm diag}\left({\bf a}\right)$ is a diagonal matrix whose $i$-th diagonal element is $\left[{\bf a}\right]_{i}$, and $\|{\bf a}\|_p$ is the $l_p$-norm of ${\bf a}$.
$\left[{\bf A}\right]_{{\cal I},:}$ denotes the sub-matrix consisting of the rows of $\bf{A}$ indexed by the ordered set $\cal I$.
${{{\left\langle k  \right\rangle }_N}}$ is the remainder of $k$ after divided by $N$.
${\rm{rect}}\left( t \right) \buildrel \Delta \over = 1$ for $t \in \left[ {0,1} \right)$ and $0$ otherwise.
$\delta \left( x \right)$ is the Dirac delta function.
${\cal CN}\left(\mu,\sigma^2 \right)$ denotes the complex Gaussian distribution with mean $\mu$ and variance $\sigma^2$, while ${\cal U}\left(a,b \right)$ denotes the uniform distribution within $\left( a,b \right)$.
$\otimes$ and $\odot$ stand for Kronecker product and Khatri-Rao product, respectively, and ${\rm vec}(\cdot )$ vectorizes a matrix by stacking its columns.

%% file: Sec/Sec2_Preliminaries.tex
\section{OCDM Preliminaries}\label{sec:2}

\begin{figure}[!tp]
\captionsetup{font={footnotesize}, name = {Fig.}, singlelinecheck=off, labelsep = period}
\begin{center}
\includegraphics[width=3.4in]{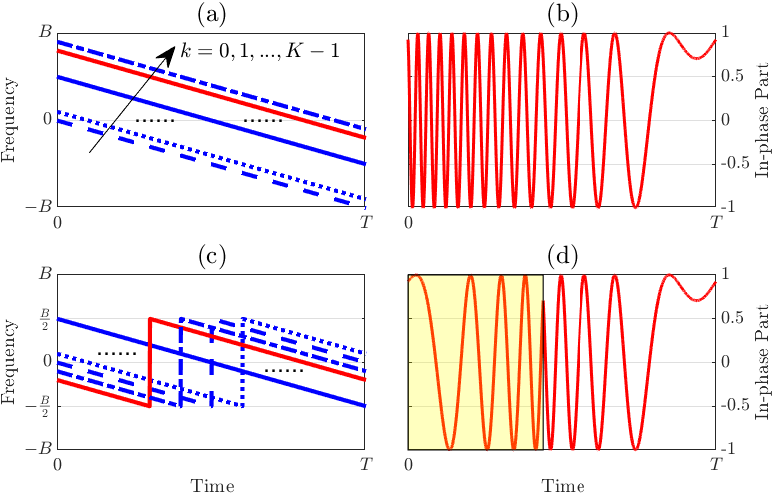}
\end{center}
\caption{OCDM signal representations. (a) The instantaneous frequencies. (b) The time-domain representation of one subcarrier that marked by red in (a). (c) The instantaneous frequencies with the folded spectrum. (d) The spectrum-folded version of one subcarrier that marked by red in (c).}
\label{fig:TFsignals} 
\end{figure}

The OCDM is based on the fact that a set of $K$ chirp signals $\left\{ {{\psi _k}\left( t \right)} \right\}_{k = 0}^{K - 1}$ with
\begin{align}\label{equ:sc} 
  {\psi _k}\left( t \right) = {\rm{rect}}\left( t/T \right){e^{\textsf{j}\frac{\pi }{4}}}{e^{ - \textsf{j}\pi \frac{K}{T^2}{{\left( {t - k\frac{T}{K}} \right)}^2}}}
\end{align}
exhibit mutual orthogonality{\footnote[1]{Equation~\eqref{equ:sc} assumes that $K$ is even. The expression of ${\psi _k}\left( t \right)$ when $K$ is odd can be found in \cite{ouyang2016orthogonal}.}}, i.e., $ \int_0^T {\psi _{k'}^*\left( t \right){\psi _k}\left( t \right)} {\rm d}t = 0$ for $k \ne k'$, where $T$ is the duration of each chirp signal and, accordingly, the bandwidth of each chirp signal is $B = K/T$. 
The instantaneous frequencies and the time-domain representations of ${\psi _k}\left( t \right)$ are presented in Figs.~\ref{fig:TFsignals}\,(a) and (b), respectively.
Given the mutual orthogonality, this set of chirp signals can be used as multiple orthogonal carriers to transmit at most $K$ data streams without cross-interference, instead of using sinusoidal carriers as in OFDM systems. 
However, given that each OCDM subcarrier is a frequency-shifted version of a ``root'' chirp with bandwidth $B$ \cite{ouyang2016orthogonal,omar2021performance}, the maximum bandwidth of $K$ OCDM subcarriers can be up to $2B$ (see Fig.~\ref{fig:TFsignals}\,(a)). 
To fit $K$ OCDM subcarriers within a bandwith of $B$, the OCDM waveform can be discretized by
collecting $K$ samples of $\psi_k\left( t \right)$ in \eqref{equ:sc} with the Nyquist sampling period $T_{\rm s} = 1/B$, and then applying a pulse-shaping filter spanning frequencies between $-B/2$ and $B/2$. This results in the following spectrum-folded OCDM waveform
\begin{align}\label{equ:sc_wrapped} 
  {{\overline \psi }_k}\left( t \right) = {\rm{rect}}\left( t/T \right){e^{\textsf{j}\frac{\pi }{4}}}{e^{\textsf{j}2\pi \left( {\int_0^t {\phi _k^{}\left( x \right){\rm{d}}x}  - \frac{{{k^2}}}{{2K}}} \right)}},
\end{align}
where $\phi _k^{}\left( x \right) \buildrel \Delta \over = {\left\langle { - B{{x}}/{{{T}}} + {{k}/{T} + B/2} } \right\rangle _B} - {B}/{2}$. Examples of ${{\overline \psi }_k}\left( t \right) $ are provided in Figs.~\ref{fig:TFsignals}(c) and (d).
Equation~\eqref{equ:sc_wrapped} can be further simplified by permuting the indices of the OCDM subcarriers, i.e.,
$\left\{ {{\widetilde \psi _k}\left( t \right)} =  {{\overline \psi _{\left\langle k-K/2 \right\rangle_K}}\left( t \right)} \right\}_{k = 0}^{K - 1}$. This leads to
\begin{align}\label{equ:sc_permuted} 
  {{\widetilde \psi }_k}\left( t \right) = {\rm{rect}}\left( t/T \right){e^{\textsf{j}\frac{\pi }{4}}}{e^{\textsf{j}2\pi \left( {\int_0^t {\widetilde \phi _k^{}\left( x \right){\rm{d}}x}  - \frac{{{{\left\langle k-K/2 \right\rangle^2_K}}}}{{2K}}} \right)}},
\end{align}
where $\widetilde \phi _k^{}\left( x \right) \buildrel \Delta \over = {\left\langle { - B{{x}}/{{{T}}} +  {{k}/{T}}} \right\rangle _B} - {B}/{2}$.

Although OCDM has been considered as a waveform candidate for communications \cite{ouyang2016orthogonal,omar2021performance,ouyang2023channel,zhang2021channel,wang2023underwater,jia2024ocdm} and ISAC \cite{bhattacharjee2022evaluation,li2024orthogonal,li2023thz,de2021joint}, the impact of spectrum-folded property has rarely been investigated.
However, it has a significant influence on the sensing performance of OCDM with analog processing, as explained in Section \ref{sec:4}.

%% file: Sec/Sec3_Model.tex
\section{System Description and Channel Model}\label{sec:3}

\begin{figure}[tp!]
\captionsetup{font={footnotesize,color={black}}, name = {Fig.}, singlelinecheck=off, labelsep = period}
\begin{center}
\includegraphics[width=3.4in]{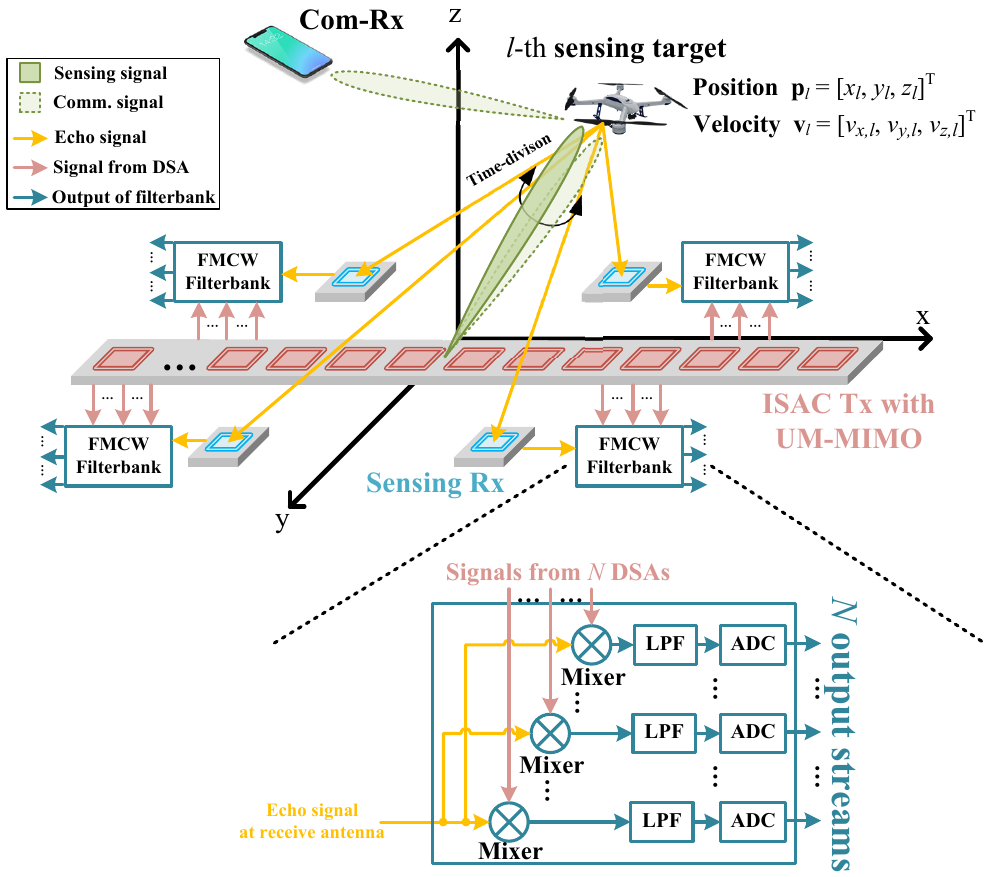}
\end{center}
\caption{The proposed ISAC scheme with UM-MIMO and OCDM-FMCW paradigm.}
\label{fig:Model} 
\end{figure}

This section presents our proposed ISAC system based on the emerging UM-MIMO and OCDM, which facilitates both near-field sensing and communications. 

\subsection{Proposed ISAC Architecture}\label{S3.1}

The architecture of the proposed ISAC station assisted by UM-MIMO and OCDM is depicted in Fig.~\ref{fig:Model}. It consists of an ISAC transmitter with $N_{\rm Tx}$ transmit (Tx) antennas and a co-located sensing receiver with $N_{\rm Rx}$ receive (Rx) antennas.
The carrier frequency of the system is $f_{\rm c}$.
We consider the UM-MIMO at the ISAC transmitter with a large $N_{\rm Tx}$ (e.g., $N_{\rm Tx} = 512$), while a few sensing Rx antennas (e.g., $N_{\rm Rx} = 4$) are sparsely located to form a large effective aperture \cite{gao2022integrated}.
For ease of analysis, we use the three-dimensional Cartesian coordinate system as shown in Fig.~\ref{fig:Model}, and assume that all the antennas at the ISAC station are located on the xy-plane. The coordinates of the $i$-th Tx antenna and $j$-th Rx antenna are respectively denoted by ${\bf p}_i^{\rm Tx} \in \mathbb{R}^3$, $i = 1,2,\ldots ,N_{\rm Tx}$, and
${\bf p}_j^{\rm Rx} \in \mathbb{R}^3$, $j = 1,2,\ldots ,N_{\rm Rx}$.
In the sensing receiver, each Rx antenna is equipped with a FMCW filterbank having $N < \min \{K, N_{\rm Tx}\}$ output ports, as shown in Fig.~\ref{fig:Model}.
The FMCW filterbank comprises $N$ instances of FMCW receiver \cite{wan2024orthogonal,patole2017automotive}, each incorporating a mixer, an analog low-pass filter (LPF), and a low-rate ADC.

\begin{figure}[t]
\captionsetup{font={footnotesize}, name = {Fig.}, singlelinecheck=off, labelsep = period}
\begin{center}
\includegraphics[width=0.48\textwidth]{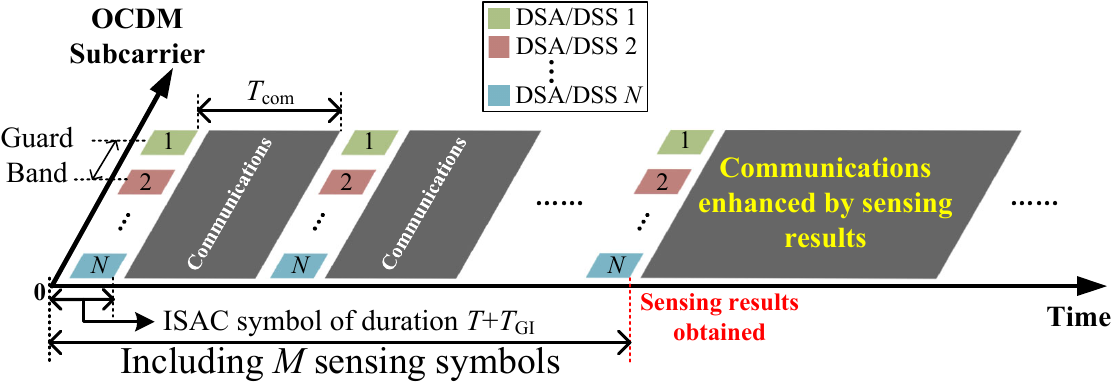}
\end{center}
\caption{The frame structure of the proposed ISAC scheme.}
\label{fig:FS} 
\end{figure}

{\color{red}
In Fig.~\ref{fig:FS}, we illustrate the transmission frame structure of the proposed ISAC system, where the ISAC transmitter sends the OCDM-based waveform either for communications or sensing in a time-division manner, which provides critical physical advantages.
First, the temporal isolation fundamentally decouples the frequency-domain resources, allowing the sensing tasks to utilize the entire system bandwidth $B$ to achieve the maximum physical range resolution ($\Delta r = c/2B$) without competing with communication subcarriers.
Second, it circumvents the self-interference inherent in full-duplex operation.
Third, the scheme optimizes the trade-off between Doppler sensitivity and resource efficiency. Since high-resolution velocity estimation requires long observation intervals to accumulate Doppler effects, we interleave communication symbols between these extended sensing symbols.
This ensures that the time resources are fully exploited for high-rate data transmission while simultaneously satisfying the requirements for precise velocity estimation.
}

\subsection{Near-Field Channel Model for UM-MIMO}\label{S3.2}

This subsection presents the applied channel models with UM-MIMO for communications and sensing respectively.

\subsubsection{Sensing Channel Model}
 
Given $L$ near-field sensing targets, the position and the velocity of the $l$-th target are respectively denoted as ${{\bf{p}}_l} = {\left[ {{x_l},{y_l},{z_l}} \right]^{\rm{T}}} \in \mathbb{R}^3$ ($z_l > 0$) and ${{\bf{v}}_l} = {\left[ {{v_{x,l}},{v_{y,l}},{v_{z,l}}} \right]^{\rm{T}}} \in \mathbb{R}^3$, $l = 1,2,\ldots ,L$.
The sensing channel from the $i$-th Tx antenna to the $j$-th Rx antenna associated with the $l$-th target can be represented as the following time-variant system
\begin{align}\label{equ:CM} 
  {h_{i,j,l}}\left( {t,\tau } \right) = {\alpha _{i,j,l}}\delta \left( {\tau  - {{d_{i,j,l}}\left( t \right)}/{c}} \right),
\end{align}
where $c$ is the speed of light, ${\alpha _{i,j,l}}$ is the complex gain, and
\begin{align}\label{eqDist} 
    {d_{i,j,l}}\left( t \right) = \left\| {{{\bf{p}}_l} + {{\bf{v}}_l}t - {\bf{p}}_i^{{\rm{Tx}}}} \right\|_2^{} + \left\| {{{\bf{p}}_l} + {{\bf{v}}_l}t - {\bf{p}}_j^{{\rm{Rx}}}} \right\|_2^{}
\end{align}
is the distance of the propagation path as the function of time.

\begin{figure}[tp!]
\captionsetup{font={footnotesize,color={red}}, name = {Fig.}, singlelinecheck=off, labelsep = period}
\begin{center}
\subfigure[]{\includegraphics[width=0.15\textwidth]{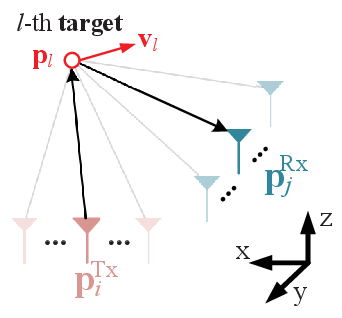}}
\subfigure[]{\includegraphics[width=0.15\textwidth]{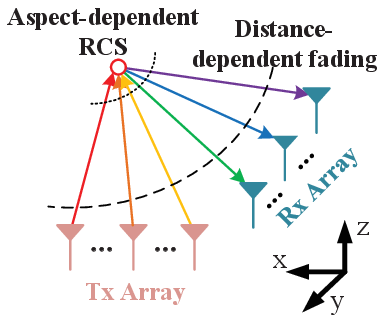}}
\subfigure[]{\includegraphics[width=0.15\textwidth]{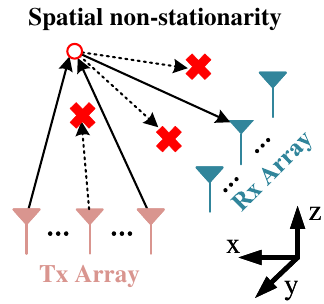}}
\end{center}
\caption{The adopted channel models: (a) The pairwise channel model. (b) Spatial uncorrelated (SUC) channel. (c) Spatial non-stationarity (SNS).}
\label{fig:CM}
\end{figure}

Note that most of the existing works on UM-MIMO communications \cite{cui2022channel} and ISAC \cite{wang2024performance,wang2024cramer} heavily depend on the assumption that the channel is spatially-correlated, i.e., $\alpha_{i,j,l} = \alpha_{l}$, $\forall i,j$ {\color{red}(see Fig.~\ref{fig:CM}(a)).
However, this assumption becomes impractical in near-field communications/sensing for the following reasons: (i) Due to its large aperture, the UM-MIMO sees each near-field target via large angle spread and delay spread, which makes radar cross section (RCS) and fading aspect-dependent and distance-dependent, respectively, as seen in Fig.~\ref{fig:CM}(b). This enhances the distinctiveness of complex gain for each Tx-Rx antenna pairs, which in turn increases the overall channel diversity. We refer to this as a {\em spatially-uncorrelated} (SUC) channel; and (ii) the near-field target may only see a portion of the UM-MIMO, as indicated in Fig.~\ref{fig:CM}(c), particularly in the indoor scenarios, which renders $\alpha_{i,j,l} = 0$ for some $i,j,l$. This is called {\em spatial non-stationarity} (SNS) in the context of near-field communications \cite{de2020non}, but it is rarely considered in the sensing scenarios.
}

\subsubsection{Communication Channel Model}

Since small timescale is considered for communication tasks \cite{gao2022integrated}, we formulate the communication channel as the time-invariant systems.
We consider one single-antenna user terminal (UT) and $L^{\rm com}$ scatterers both located in the near-field regime of the ISAC station.
The positions of UT and $l$-th scatterer are respectively denoted as 
${\bf p}^{\rm UT} \in \mathbb{R}^3$ and ${{\bf{p}}^{\rm com}_l} \in \mathbb{R}^3$, ${l} = 1,2,\ldots,L^{\rm com}$. The communication channel impulse response (CIR) from the ISAC station to the UT associated can be represented as
\begin{align}\label{CMcom} 
  {{\bf h}^{\rm com}}\left( {\tau } \right) = \sum\nolimits_{l = 1}^{{L^{{\rm{com}}}}} {\alpha _l^{{\rm{com}}}{\bf{a}}_l^{{\rm{com}}}p_{\rm PSF}\left( {\tau  - \tau _l^{{\rm{com}}}} \right)} ,
\end{align}
where ${\alpha^{\rm com}_{{l}}}$ is the complex gain, $p_{\rm PSF}\left( \cdot \right)$ is the pulse shaping filter function, $\tau_l^{\rm com} = \left\| {{{\bf{p}}^{\rm com}_{l}}} \right\|_2^{} + \left\| {{{\bf{p}}^{\rm com}_{l}} - {\bf{p}}^{{\rm{UT}}}} \right\|_2^{}/c$, and ${\bf{a}}_l^{{\rm{com}}} \in \mathbb{C}^{N_{\rm Tx}}$ is the near-field steering vector associated with the $l$-th scatter. The $i$-th element ($i = 1,2,\ldots,N_{\rm Tx}$) of ${\bf{a}}_l^{{\rm{com}}}$ is given by \cite{cui2022channel}
\begin{align}\label{NFSV} 
    \left[{{{\bf a}^{\rm com}_{{l}}}}\right]_i = \frac{1}{\sqrt{N_{\rm Tx}}}e^{-\textsf{j}{2\pi}{f_{\rm c}}(\left\| {{{\bf{p}}^{\rm com}_{l}} - {\bf{p}}_i^{{\rm{Tx}}}} \right\|_2^{} - \left\| {{{\bf{p}}^{\rm com}_{l}}} \right\|_2)}.
\end{align}
Note that since the single-antenna UT is considered, the UT sees each scatterer via a small angle spread, so the spatially-correlated channel model is adopted for communication tasks in \eqref{CMcom}, i.e., the complex gain is the same for each scatterer.

%% file: Sec/Sec4_Sensing.tex
\section{Proposed Near-Field Sensing Paradigm}\label{sec:4}

In this section, we propose the near-field sensing paradigm aided by UM-MIMO and OCDM techniques.
	
    \begin{figure*}\setcounter{equation}{14}  
    \begin{align}\label{equ:IF} 
        s^{\rm IF}_{n,n'} \left( t ; \tau \right) = e^{\textsf{j}\varphi _{n,n'}} \times \left\{ {\begin{array}{*{20}{l}}
            {{e^{\textsf{j} \varphi _{n}^{\rm{I}}\left( \tau  \right)}}\exp \left( {\textsf{j}2\pi \left( {\frac{B\tau}{T} + \frac{{{k_{n'}} - {k_{n}}}}{T}} \right)t} \right),}&{t \in {\cal T}_{n,n'}^{\rm{I}}\left( \tau  \right)} , \\
            {{e^{\textsf{j} \varphi _{n,n'}^{{\rm{II}}}\left( \tau  \right)}}\exp \left( {\textsf{j}2\pi \left( {\frac{B\tau}{T} + \frac{{{k_{n'}} - {k_{n}}}}{T} + B{\Delta _{n,n'}}\left( \tau  \right)} \right)t} \right),}&{t \in {\cal T}_{n,n'}^{\rm{II}}\left( \tau  \right)} , \\
            {{e^{\textsf{j} \left( \varphi _{n}^{{\rm{I}}}\left( \tau  \right) - 2\pi B \tau \right) }}\exp \left( {\textsf{j}2\pi \left( {\frac{B\tau}{T} + \frac{{{k_{n'}} - {k_{n}}}}{T}} \right)t} \right),}&{t \in {\cal T}_{n,n'}^{\rm{III}}\left( \tau  \right)} ,
        \end{array}} \right.
    \end{align}
    \hrule
    \end{figure*}
		
\subsection{OCDM-based Sensing Waveform}\label{S4.1}

Since we focus on the analog processing with low-sampling-rate ADCs for sensing tasks, the sensing problem formulation in this section will be based on the continuous-time representation of OCDM waveform \eqref{equ:sc_permuted}.
As indicated by Fig.~\ref{fig:FS}, the ISAC transmitter will transmit designed OCDM symbols one by one. We propose that each sensing symbol consists of only $N$ subcarriers, namely, $N$ DSSs, and each DSS is transmitted by only one Tx antenna, call a DSA (and other antennas will be shut down for the moment). In other words, there is a one-to-one mapping between $N$ DSSs (out of $K$ OCDM subcarriers) and $N$ DSAs (out of $N_{\rm Tx}$ Tx antennas).
We introduce the ordered sets ${\cal I}_{\rm DSS} = \{ k_1,k_2,\ldots,k_N \} \subseteq \{ 0,1,\ldots,K-1 \}$ and ${\cal I}_{\rm DSA} = \{ i_1,i_2,\ldots,i_N \} \subseteq \{ 1,2,\ldots,N_{\rm Tx}\}$ to represent the selection schemes of DSS and DSA, respectively.
Considering that $M$ ISAC symbols are transmitted, the waveform at the $n$-th DSA of the $m$-th ISAC symbol can be written as\setcounter{equation}{7}
\begin{align}\label{equ:smn} 
  {s_{m,n}}\left( t \right) = \sqrt {{{{P_{{\rm{Tx}}}}}}/{{N}}} {e^{\textsf{j}2\pi {f_{\rm{c}}}\left( {t - T_m} \right)}} {\widetilde \psi _{k_n}} \left( {t - T_m} \right),
\end{align}
where $m = 1,2,\ldots,M$, $n = 1,2,\ldots,N$, $k_n \in {\cal I}_{\rm DSS}$, $T_m = \left(m-1\right)\left( T+T_{\rm GI}+{T_{{\rm{com}}}} \right) + {T_{{\rm{GI}}}}$, $T_{\rm com}$ is the duration of the communication symbol between two adjacent sensing symbols, and $P_{\rm Tx}$ is the transmit power of ISAC symbols. Unlike the CP for communications, the zero padding (ZP) is adopted as the GI before each sensing symbol, as it can not only eliminate the ISI, but also preserve the idle time for reconfiguring the RF circuits \cite{gao2022integrated}.

\subsection{OCDM Meets FMCW}\label{S4.2}

For brevity, we define ${{\overline \alpha }_{n,j,l}} = {\alpha _{{i_n},j,l}}$, ${{\overline h }_{n,j,l}} = {h _{{i_n},j,l}}$, and ${{\overline d}_{n,j,l}} \left( t \right) = {d_{{i_n},j,l}} \left( t \right)$ where $i_n \in {\cal I}_{\rm DSA}$.
The noise at the sensing receiver is temporally ignored. Based on the transmit waveform \eqref{equ:smn} and the sensing channel \eqref{equ:CM}, the received echo signal of the $m$-th sensing symbol at the $j$-th Rx antenna can be expressed as
\begin{align}\label{equ:rjm} 
  {r_{j,m}}\left( t \right) & = \sum\limits_{n' = 1}^N {\sum\limits_{l = 1}^L {\int_0^{ + \infty } {{{\overline h}_{{n'},j,l}}\left( {t,\tau } \right)} {s_{m,n'}}\left( {t - \tau } \right){\rm{d}}\tau } } \nonumber \\
& = \sum\limits_{n' = 1}^N {\sum\limits_{l = 1}^L {{{\overline \alpha} _{n',j,l}}{s_{m,n'}}\left( {t - {{{{\overline d}_{n',j,l}}\left( t \right)}}/{c}} \right)} } ,
\end{align}
for $t \in \left[ {T_m,T_m + T} \right)$. 
Note that the length of GI should satisfy ${T_{{\rm{GI}}}} > \mathop {\max }\limits_{n,j,l,t} \left\{ {{{\overline d}_{n,j,l}\left(t\right)}} \right\}/c$ throughout the sensing stage to guarantee the ISI-free form \eqref{equ:rjm}.
The output of the mixer connecting the $n$-th DSA and the $j$-th Rx antenna is \cite{patole2017automotive,wan2024orthogonal}
\begin{align}\label{equ:mixer} 
  r_{n,j,m}^{{\rm{Mix}}}\left( t \right) =&  {r_{j,m}}\left( t \right)s_{m,n}^ * \left( t \right) = \frac{{{P_{{\rm{Tx}}}}}}{{N}} \sum\limits_{n' = 1}^N \sum\limits_{l = 1}^L {{\overline \alpha} _{n',j,l}} \nonumber \\
	& \hspace*{-10mm}\times {e^{ - \textsf{j}2\pi \frac{f_{\rm{c}}}{c} {{{{\overline d}_{n',j,l}}\left( t \right)}}}} s_{n,n'}^{{\rm{IF}}}\left( t -T_m ; {{{{\overline d}_{n',j,l}}\left( t \right)}}/{c} \right),
\end{align}
where the IF signal obtained by mixing two DSSs in the OCDM system is
\begin{align}\label{equ:DefinesIF} 
  s_{n,n'}^{{\rm{IF}}}\left( t ; \tau \right) =  {{\widetilde \psi} _{{k_{n'}}}}\left( {t - \tau} \right){\widetilde \psi} _{{k_{n}}}^ * \left( {t} \right)
\end{align}
with $k_n,k_{n'} \in {\cal I}_{\rm DSS}$. Furthermore, we can simplify \eqref{equ:mixer} as
\begin{align}\label{equ:mixerApprox} 
    r_{n,j,m}^{{\rm{Mix}}}\left( t \right) & \approx \frac{{{P_{{\rm{Tx}}}}}}{{N}}\sum\limits_{n' = 1}^N {\sum\limits_{l = 1}^L {{{\overline \alpha}'_{n',j,l}}{e^{ - \textsf{j}2\pi \frac{{v_{n',j,l}}}{\lambda}  t }}} } \nonumber \\ 
    & \qquad \qquad \quad 
    \times s_{n,n'}^{{\rm{IF}}}\left( t -T_m ; {{{{\tau}_{n',j,l}}}} \right) ,
\end{align}
where $\lambda = c/f_c$ is the wavelength.
The approximation \eqref{equ:mixerApprox} is due to the following reasons: (i) we assume the positions of targets to be viewed as invariant during the sensing stage, and therefore replace ${{{{\overline d}_{n,j,l}}\left( t \right)}}/{c}$ by
\begin{align}\label{equ:DisDefine} 
  {{{{\tau}_{n,j,l}}}} = {{{{\overline d}_{n,j,l}}\left( 0 \right)}}/{c} = \frac{{{{\left\| {{{\bf{p}}_l} - {\bf{p}}_{{i_n}}^{{\rm{Tx}}}} \right\|}_2} + {{\left\| {{{\bf{p}}_l} - {\bf{p}}_j^{{\rm{Rx}}}} \right\|}_2}}}{c}
\end{align}
in the formulation of IF signal; and (ii) we use the Taylor approximation ${{\overline d}_{n,j,l}}\left( t \right) \approx {{\overline d}_{n,j,l}}\left( 0 \right) + v_{n,j,l}t$ with 
\begin{align}\label{equ:VelDefine} 
  {v_{n,j,l}} & =  {\left. {\frac{{\partial {{\overline d}_{n,j,l}}\left( t \right)}}{{\partial t}}} \right|_{t = 0}} \nonumber \\
   & = {\left( {\frac{{{{\bf{p}}_l} - {\bf{p}}_{{i_n}}^{{\rm{Tx}}}}}{{\left\| {{{\bf{p}}_l} - {\bf{p}}_{{i_n}}^{{\rm{Tx}}}} \right\|_2^{}}} + \frac{{{{\bf{p}}_l} - {\bf{p}}_j^{{\rm{Rx}}}}}{{\left\| {{{\bf{p}}_l} - {\bf{p}}_j^{{\rm{Rx}}}} \right\|_2^{}}}} \right)^{\rm{T}}}{{\bf{v}}_l}
\end{align}
to modify the phase term in \eqref{equ:mixer}, and therefore to obtain \eqref{equ:mixerApprox} where ${\overline \alpha}'_{n,j,l} = {\overline \alpha}_{n,j,l} e^{-\textsf{j} 2\pi f_{\rm c} \tau_{n,j,l}}$.

In conventional FMCW radar systems (i.e., $k_n = 0$, $\forall n$), the IF signal manifests as a single-tone sinusoidal waveform whose frequency (i.e., beat frequency) is mainly determined by the delay of the received signal. However, when spectrum-folded OCDM subcarriers \eqref{equ:sc_permuted} are employed for sensing, the characteristic of the IF signal will be distinct, as elaborated in the following theorem.

\begin{theorem}[The IF signal expression]\label{Theorem1}
    Based on \eqref{equ:DefinesIF}, the IF signal $s_{n,n'}^{{\rm{IF}}}\left( t ; \tau \right)$ can be expressed as \eqref{equ:IF} shown at the top of the page, where ${{\cal T}_{n,n'}^{\rm{I}}\left( \tau  \right) = \left[ {\tau, \, T_{n,n'}^{{\rm{BWS}}}\left( \tau  \right)} \right)}$, ${\cal T}_{n,n'}^{\rm{II}}\left( \tau  \right) = {\left[ {{ T}_{n,n'}^{{\rm{BWS}}}\left( \tau  \right), \, T_{n,n'}^{{\rm{BWE}}}\left( \tau  \right)} \right)}$, ${{\cal T}_{n,n'}^{{\rm{III}}}\left( \tau  \right) = \left[ {T_{n,n'}^{{\rm{BWE}}}\left( \tau  \right),\, T} \right)}$, and\setcounter{equation}{15}
    \begin{align} 
        T_{n,n'}^{{\rm{BWS}}}\left( \tau  \right) & = \max \left\{ {\tau,\min \left\{ {{k_n}/B,{k_{n'}}/B + \tau } \right\}} \right\}, \\
        T_{n,n'}^{{\rm{BWE}}}\left( \tau  \right) & = \min \left\{ {T,\max \left\{ {{k_n}/B,{k_{n'}}/B + \tau } \right\}} \right\}.
    \end{align}
    The expressions of $\varphi _{n,n'}$, $\varphi _{n}^{\rm{I}}\left( \tau  \right)$, $\Delta_{n,n'}\left( \tau  \right)$, and $\varphi _{n,n'}^{\rm{II}}\left( \tau  \right)$ are:
    \begin{align} 
        \label{equ:A2}
        {\varphi _{n,n'}} & = \frac{\pi }{K}\left( {\left\langle {k_n - K/2} \right\rangle _K^2 - \left\langle {k_{n'} - K/2} \right\rangle _K^2} \right), \\
        \label{equ:A4}
        \varphi _n^{\rm{I}}\left( \tau  \right) & = - \frac{B\pi}{{T}}{\tau ^2} - \left( {\frac{{{2\pi k_n}}}{T} - {B\pi}} \right)\tau, \\
        \label{equ:A8}
        \Delta_{n,n'}\left( \tau \right) & =  \left\{ {\begin{array}{*{20}{l}}
        { - 1,}&{{k_n} - {k_{n'}} < B\tau ,}\\
        {1,}&{{k_n} - {k_{n'}} \ge B\tau ,}
        \end{array}} \right. \\
        \label{equ:A9}
        \varphi _{n,n'}^{{\rm{II}}}\left( \tau  \right) & = \varphi _n^{\rm{I}}\left( \tau  \right) - \pi \left[ {{1 + {\Delta _{n,n'}}\left( \tau  \right)}} \right] B\tau .
    \end{align}
\end{theorem}
{\color{red}
\begin{proof}
    Refer to {Appendix~\ref{app:a}}.
\end{proof}
}

\begin{figure}[!tp]
\captionsetup{font={footnotesize}, name = {Fig.}, singlelinecheck=off, labelsep = period}
\centering
\includegraphics[width=0.4\textwidth]{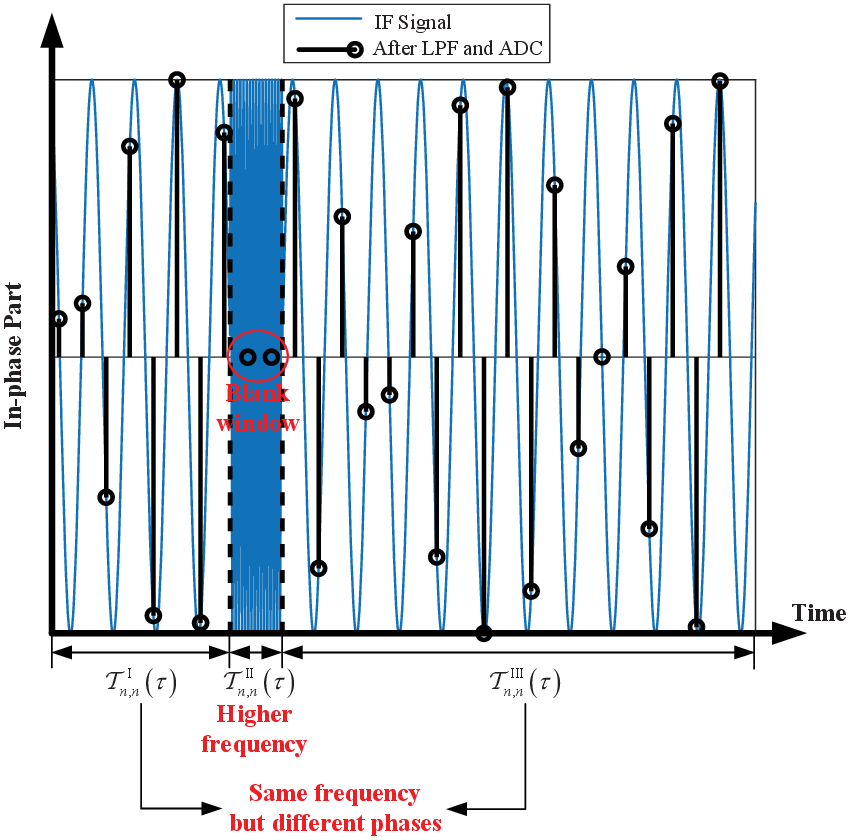}
\caption{An example of the IF signal (in-phase part) when $n'=n$ in the OCDM system and its corresponding digitization.}
\label{fig:sIF} 
\end{figure}

We provide an example of $s^{\rm IF}_{n,n'} \left( t ; \tau \right)$ for $n'=n$ in Fig.~\ref{fig:sIF}. 
Our objective is to preserve the desired beat frequency ${\frac{B\tau}{T}}$ (for $n'=n$) and eliminate all other frequency components, which can be achieved by judiciously designing ${\cal I}_{\rm DSS}$ and the cut-off frequency of the LPF. Assuming that the LPF only preserves the frequency components between $-f_{\rm LPF}$ and $f_{\rm LPF}$, the criterion for designing ${\cal I}_{\rm DSS}$ and $f_{\rm LPF}$ is summarized as
\begin{align}\label{equ:criterion} 
  \left\{ {\begin{array}{*{20}{l}}
  {\Big| {\frac{{B\tau }}{T}} \Big| < {f_{{\rm{LPF}}}} ,} \vspace*{1mm}\\
  {\Big| {\frac{{B\tau }}{T} + \frac{{{k_{n'}} - {k_n}}}{T}} \Big| > {f_{{\rm{LPF}}}} ,} \vspace*{1mm}\\
  {\Big| {\frac{{B\tau }}{T} + \frac{{{k_{n'}} - {k_n}}}{T} \pm B} \Big| > {f_{{\rm{LPF}}}} ,}
  \end{array}} \right. \forall \tau, n' \neq n .
\end{align}


Given that $\tau < T_{\rm GI}$ and applying the triangle inequality $\left| a \right| - \left| b \right| \le \left| {a \pm b} \right| \le \left| a \right| + \left| b \right|$, $\forall a,b$, one can readily obtain a solution that fulfills \eqref{equ:criterion} as
\begin{align}\label{equ:solution} 
  \left\{ {\begin{array}{{l}}
  {{f_{{\rm{LPF}}}} \ge B{T_{{\rm{GI}}}}/T ,}\\
  {\left| {{k_{n'}} - {k_n}} \right| \ge B{T_{{\rm{GI}}}} + {f_{{\rm{LPF}}}}T ,}\\
  {\left| {{k_{n'}} - {k_n}} \right| \le K - B{T_{{\rm{GI}}}} - {f_{{\rm{LPF}}}}T ,}
  \end{array}} \right. \forall n' \neq n.
\end{align}

In a nutshell, as long as ${\cal I}_{\rm DSS}$ and the cut-off frequency of LPF $f_{\rm LPF}$ are designed to fulfill the conditions in \eqref{equ:solution}, the LPF can preserve the desired IF components for delay estimation while eliminating the interference from all other DSSs. Accordingly, the output of the LPF would be
\begin{align}\label{equ:LPF} 
  r_{n,j,m}^{{\rm{LPF}}}\! \left( t \right)\! =\! \frac{{{P_{{\rm{Tx}}}}}}{{N}}\! \sum\limits_{l = 1}^L\! {{{\overline \alpha}'_{n,j,l}} {e^{ - \textsf{j}2\pi \frac{v_{n,j,l}}{\lambda} t}}} {\overline s}_{n}^{{\rm{IF}}}\!\left( {t\! -\! {T_m};{\tau _{n,j,l}}} \right)\! ,\!
\end{align}
where only the component $n' = n$ in \eqref{equ:mixerApprox} is preserved, and
\begin{align} 
  {\overline s}_n^{{\rm{IF}}}\left( {t;\tau } \right){\rm{ }} = \left\{ 
  {\begin{array}{*{20}{l}}
  {0,}&{t \in {\cal T}_{n,n}^{\rm{II}}\left( \tau  \right) ,}\\
  {s_{n,n}^{{\rm{IF}}}\left( {t;\tau } \right),}&{{\rm{otherwise}} .}
  \end{array}} \right. 
\end{align}
In this way, we have decoupled the received signals from different antenna pairs $\left(n,j\right)$, $n = 1,2,\ldots,N$, $j = 1,2,\ldots,N_{\rm Rx}$, and thus obtained $NN_{\rm Rx}$ independent sensing measurements for each sensing symbol. 

{\color{red}
\begin{remark}\label{REM2}
This independence among different DSSs/DSAs makes the sensing process immune to the SUC channels and SNS as depicted in Fig.~\ref{fig:CM}.
More importantly, the bandwidth of the LPF output will be confined to $f_{\rm LPF}$, which can be much smaller than the system bandwidth $B$. This allows us to use the low-rate ADC to sample $r_{n,j,m}^{{\rm{LPF}}}\left( t \right)$ for digital processing, significantly alleviating the burden of hardware complexity and power consumption compared to the existing OFDM-based \cite{wang2024performance} and OTFS-based \cite{gaudio2020effectiveness} ISAC schemes. 
\end{remark}
}


We denote the ADC sampling rate as $f_{\rm ADC} \ge f_{\rm LPF}$ and let $Q = \left\lfloor {{f_{{\rm{ADC}}}}\left( {T - {T_{{\rm{GI}}}}} \right)} \right\rfloor $.
Note that $f_{\rm ADC} \ll B$.
Sampling $r_{n,j,m}^{{\rm{LPF}}}\left( t \right)$ at time $t = T_m + T_{\rm GI} + \frac{q-1}{f_{{\rm{ADC}}}}$ for $q = 1,2,\ldots,Q$ yields the measurement vector ${\bf r}_{n,j,m} \in \mathbb{C}^{Q}$ which is given by
\begin{align}\label{equ:rnjm} 
  {\bf r}_{n,j,m} & = {\left[ {r_{n,j,m}^{{\rm{ADC}}}\left[ 1 \right],r_{n,j,m}^{{\rm{ADC}}}\left[ 2 \right],\ldots ,r_{n,j,m}^{{\rm{ADC}}}\left[ {Q} \right]} \right]^{\rm{T}}} \nonumber \\
   & \approx \frac{{{P_{{\rm{Tx}}}}}}{{N}}\sum\limits_{l = 1}^L {{{\overline \alpha}''_{n,j,l}}{e^{ - \textsf{j}2\pi {\nu_{n,j,l}}\left(m-1\right)}}} {\bf{b}}_{n}\left(\mu_{n,j,l} \right),
\end{align}
where $r_{n,j,m}^{{\rm{ADC}}}\left[ q \right] = r_{n,j,m}^{{\rm{LPF}}}\left( {{T_m} + {T_{{\rm{GI}}}} + \frac{q-1}{f_{{\rm{ADC}}}}} \right)$, ${{\overline \alpha}''_{n,j,l}} = {{\overline \alpha}'_{n,j,l}} {\exp \left( {\textsf{j}2\pi \left( {\frac{{B{\tau _{n,j,l}}}}{T} - \frac{{{v_{n,j,l}}}}{\lambda }} \right){T_{{\rm{GI}}}}} + {{\textsf{j}\varphi _n^{\rm{I}} \left( {{\tau _{n,j,l}}} \right)}} \right) }$, ${\nu _{n,j,l}} = v_{n,j,l}\left( T+T_{\rm GI}+{T_{{\rm{com}}}} \right)/\lambda$, ${\mu _{n,j,l}} = B{\tau _{n,j,l}}/\left( {{f_{{\rm{ADC}}}}T} \right)$, and ${\bf{b}}_{n}\left(\mu \right) \in \mathbb{C}^{Q}$ is the response vector of the $n$-th DSS. The $q$-th element of ${\bf{b}}_{n}\left(\mu \right)$ is given by
\begin{align}\label{equ:VecT} 
  {\left[ {\bf{b}}_{n}\left(\mu \right) \right]_{q}} = & {e^{\textsf{j}2\pi {{ {{\mu}} }}\left(q-1\right)}} \times 
  \left\{ {\begin{array}{*{20}{l}}
  {1,}&{q \in \Omega _{n}^{\rm{I}}\left( \tau \right) ,}\\
  {0,}&{q \in \Omega _{n}^{{\rm{II}}}\left( \tau \right) ,}\\
  {{e^{ - \textsf{j}\mu \phi}},}&{q \in \Omega _{n}^{{\rm{III}}}\left( \tau \right) ,}
  \end{array}} \right.
\end{align}
where $\phi = 2\pi f_{\rm ADC} T$, and
\begin{align} 
  \Omega _{n}^{{\rm{type}}} \left( \tau \right) = \left\{ {q \in \left[ 1,Q \right] \left| {{T_{{\rm{GI}}}} + \frac{{q - 1}}{{{f_{{\rm{ADC}}}}}} \in {\cal T}_{n,n}^{{\rm{type}}}\left( {{\tau}} \right)} \right.} \right\}
\end{align}
with ${\rm type} \in \{{\rm I}, {\rm II}, {\rm III} \}$. The approximation in \eqref{equ:rnjm} is made because the range-dependent component predominates during the single symbol duration, allowing the velocity-dependent component to be discarded \cite{zhang2020super}. An example of ${\bf{b}}_{n}\left(\mu \right)$ is provided in Fig.~\ref{fig:sIF}.
Moreover, we collect $\{{\bf r}_{n,j,m}\}_{m=1}^{M}$ to obtain ${{\bf{R}}_{n,j}} = \left[ {{{\bf{r}}_{n,j,1}},{{\bf{r}}_{n,j,2}},\ldots,{{\bf{r}}_{n,j,M}}} \right] \in \mathbb{C}^{Q \times M}$ as
\begin{align}\label{equ:R} 
  {{\bf{R}}_{n,j}} & = \frac{{{P_{{\rm{Tx}}}}}}{{N}}\sum\limits_{l = 1}^L {{{\overline \alpha}''_{n,j,l}}} {\bf{b}}_{n}\left(\mu_{n,j,l} \right){\bf a}^{\rm H}\left(\nu_{n,j,l} \right) \nonumber \\
  & = {\bf B}_{n} \left({\bm \mu}_{n,j} \right) {\rm diag}\left( {{{\bm \alpha }}_{n,j}} \right) {\bf A}^{\rm H} \left({\bm \nu}_{n,j} \right),
\end{align}
where ${\bf{a}}\left(\nu \right)\! =\! {\left[ {1,{e^{\textsf{j}2\pi {\nu}}},\ldots,{e^{\textsf{j}2\pi {\nu}\left( {M - 1} \right)}}} \right]^{\rm{T}}}\! \in\! \mathbb{C}^M$ is the response vector accounting for velocity,
${\bm \mu}_{n,j}\! =\! \left[ \mu _{n,j,1}, \ldots,\mu _{n,j,L} \right]^{\rm{T}}\! \in\! \mathbb{R}^L$,
${\bm \nu}_{n,j}\! =\! \left[ \nu _{n,j,1}, \ldots,\nu _{n,j,L} \right]^{\rm{T}}\! 
\in\! \mathbb{R}^L$,
${{\bm{\alpha }}_{n,j}}\! =\! \left({{{P_{{\rm{Tx}}}}}}/{{N}}\right) {\left[ {{{\overline \alpha}''_{n,j,1}},\ldots,{{\overline \alpha}''_{n,j,L}}} \right]^{\rm{T}}}\! \in\! \mathbb{C}^L$,
${{\bf{A}}}\left({\bm \nu}_{n,j} \right)\! =\! \left[ {{\bf{a}}\left(\nu_{n,j,1} \right),\ldots,{\bf{a}}}\left(\nu_{n,j,L} \right) \right]\! \in\! \mathbb{C}^{M \times L}$,
and ${{\bf{B}}_{n}} \left({\bm \mu}_{n,j} \right)\! =\! \left[ {{\bf{b}}_{n}\left(\mu_{n,j,1} \right),\ldots,{\bf{b}}_{n}\left(\mu_{n,j,L} \right)} \right]\! \in\! \mathbb{C}^{Q \times L}$.

{\color{red}Note that the aforementioned signal model is obtained under noise-free scenarios. Unlike the traditional digital processing that simply formulates the noise as the additive white Gaussian noise (AWGN), the mathematical expression of the noise in the proposed processing will be transmitted-signal-dependent and thus can be quite complex. The specific statistical analysis of the noise behavior is beyond the scope of this paper, and can be left as a future research direction.
Instead, independent and identically distributed (i.i.d.) AWGN is considered for the obtained signal model after ADC for performance evaluation. The final sensing measurements can be written as
\begin{align}\label{equ:Rn} 
  {{\bf{\overline R}}_{n,j}} = {{\bf{R}}_{n,j}} + {\bf N}_{n,j},
\end{align}
where each element in ${\bf N}_{n,j} \in \mathbb{C}^{Q \times M}$ follows the i.i.d. complex Gaussian distribution ${\cal CN}\left(0,\sigma_{\rm n}^2\right)$.}

\subsection{Parameter Estimation Under Dual Discontinuity}\label{S4.3}

\begin{algorithm}[!tp]
\caption{Proposed Target Parameter Estimation Method}
\label{ALG1}
{\bf Input}:
Received measurements ${\overline {\bf R}}_{n,j}$, ${{{\overline {\bf{r}} }_{n,j}}} = {\rm{vec}}\left( {{{\overline {\bf{R}} }_{n,j}}} \right)$, $n = 1,2,\ldots,N$, $j = 1,2,\ldots,N_{\rm Rx}$, maximum number of iterations for GDA $H_{\max}$.
\begin{algorithmic}[1]
\Statex{\it \% The algorithm will run for each antenna pair $\left(n,j\right)$}
\Statex{\it \% RL-ESPRIT module begins}
\State Estimate number of effective paths ${\widehat L}_{n,j}$;
\State Obtain ${\left[ {{{\bf{\overline R}}_{n,j}}} \right]_{\Omega _n^{{\rm{ini}}},:}}$ based on \eqref{equ:OmegaIniI}-\eqref{equ:SetIni};
\State Obtain estimates ${\widehat {\bm \mu}}_{n,j}^{{\rm{ini}}} \in \mathbb{R}^{{\widehat L}_{n,j}}$ and ${\widehat {\bm \nu}}_{n,j}^{{\rm{ini}}} \in \mathbb{R}^{{\widehat L}_{n,j}}$ via 2D-ESPRIT algorithm \cite{liao2019closed};
\State ${\widehat{\bm \alpha}}^{\rm ini}_{n,j} = \left( { {{\bf{A}}^*\left( {\widehat {\bm \nu}}_{n,j}^{{\rm{ini}}}  \right)}} \odot {{\bf{B}}_n}\left( {\widehat {\bm \mu}}_{n,j}^{{\rm{ini}}}  \right) \right)^{\dag} {{{\overline {\bf{r}} }_{n,j}}}$;
\Statex{\it \% RL-ESPRIT module ends and GDA module begins}
\State $h = 0$, ${{\bm \mu}}^{\left(0\right)} =  {\widehat {\bm \mu}}_{n,j}^{{\rm{ini}}}$, ${{\bm \nu}}^{\left(0\right)} =  {\widehat {\bm \nu}}_{n,j}^{{\rm{ini}}}$, and ${{\bm \alpha}}^{\left(0\right)} =  {\widehat {\bm \alpha}}_{n,j}^{{\rm{ini}}}$;
\While{$h < H_{\max}$,}
    \State $h = h + 1$;
    \State ${\left. {{\bf{g}}_{\rm dis} ^{\left( {h - 1} \right)} = \frac{{\partial {g_{n,j}}\left( {{\bm \mu} ,{\bm \nu} ,{\bm \alpha} } \right)}}{{\partial {\bm \mu} }}} \right|_{{\bm \mu}  = {{\bm \mu} ^{\left( {h - 1} \right)}},{\bm \nu}  = {{\bm \nu} ^{\left( {h - 1} \right)}},{\bm \alpha}  = {{\bm \alpha} ^{\left( {h - 1} \right)}}}}$,
    ${\left. \ \ \ \ {{\bf{g}}_{\rm vel} ^{\left( {h - 1} \right)} = \frac{{\partial {g_{n,j}}\left( {{\bm \mu} ,{\bm \nu} ,{\bm \alpha} } \right)}}{{\partial {\bm \nu} }}} \right|_{{\bm \mu}  = {{\bm \mu} ^{\left( {h - 1} \right)}},{\bm \nu}  = {{\bm \nu} ^{\left( {h - 1} \right)}},{\bm \alpha}  = {{\bm \alpha} ^{\left( {h - 1} \right)}}}}$;
    \State {\color{red} Choose step size for current iteration $\gamma^{\left(h\right)}$ based on Barzilai-Borwein method \cite{barzilai1988two}, as done in \eqref{equ:BB};}
    \State ${{\bm \mu} ^{\left( h \right)}} = {{\bm \mu} ^{\left( {h - 1} \right)}} - {\gamma ^{\left( h \right)}} {\bf g}_{\rm dis}^{\left(h-1\right)}$;
    \State ${{\bm \nu} ^{\left( h \right)}} = {{\bm \nu} ^{\left( {h - 1} \right)}} - {\gamma ^{\left( h \right)}} {\bf g}_{\rm vel}^{\left(h-1\right)}$;
    \State ${{\bm \alpha}}^{\left( h \right)} = \left( { {{\bf{A}}^*\left( {{\bm \nu}}^{\left( h \right)} \right)}} \odot {{\bf{B}}_n}\left( {{\bm \mu}}^{\left( h \right)}  \right) \right)^{\dag} {{{\overline {\bf{r}} }_{n,j}}}$;
\EndWhile
\end{algorithmic}
{\bf Output}: Refined estimates ${\widehat {\bm \mu}}_{n,j}^{{\rm{GDA}}} = {\bm \mu}^{\left(h\right)} \in \mathbb{R}^{{\widehat L}_{n,j}}$ and ${\widehat {\bm \nu}}_{n,j}^{{\rm{GDA}}} = {\bm \nu}^{\left(h\right)} \in \mathbb{R}^{{\widehat L}_{n,j}}$, $n = 1,2,\ldots,N$, $j = 1,2,\ldots,N_{\rm Rx}$.
\end{algorithm}

Based on ${{\bf{\overline R}}_{n,j}}$ obtained by the FMCW processing above, our objective is to estimate the parameters $\tau_{n,j,l}$ and $v_{n,j,l}$ (or $\mu_{n,j,l}$ and $\nu_{n,j,l}$ equivalently) for each antenna pair $\left(n,j\right)$. However, as shown in \eqref{equ:VecT} and Fig.~\ref{fig:sIF}, the digitized IF signals in the OCDM system may exhibit {\em dual discontinuity}. Particularly, it consists of (i) {\em time discontinuity}, which introduces the blank window (all-zero segment) to the response vectors, and (ii) {\em phase discontinuity}, which introduces an extra phase shift of $-\mu \phi$ to part of ${\bf b}_{n} \left( \mu \right)$ (see \eqref{equ:VecT}).
This destroys the Vandermonde property of ${\bf B}_{n} \left( {\bm \mu}_{n,j} \right)$ and thus invalidates many off-the-shelf parameter estimation algorithms \cite{stoica1989music,liao2019closed}.


{\color{red}It is noteworthy that while affine frequency division multiplexing (AFDM) \cite{bemani2023affine} provides a more generalized framework for multi-carrier waveforms, we specifically adopt OCDM for the proposed ISAC solution. The rationale is that the chirp subcarriers in AFDM usually exhibit steeper slopes, resulting in more than one instantaneous frequency discontinuity within the system bandwidth \cite[Fig. 4]{bemani2023affine}. Such discontinuities would significantly complicate the IF signal derivation and the subsequent super-resolution estimation. In contrast, OCDM strikes an optimal balance by providing frequency-domain diversity while maintaining the manageable discontinuities for FMCW processing.}

To tackle the dual-discontinuity, we propose a parameter estimation method taking the dual discontinuity into account, which is summarized in {Algorithm~\ref{ALG1}} and detailed as follows.

\subsubsection{Initial estimation via RL-ESPRIT}

First we estimate the number of effective paths $L_{n,j} = \left| \{ l \left| {{{\overline \alpha}''_{n,j,l}}} \neq 0 \right. \} \right|_{\rm c} \le L$ as ${\widehat L}_{n,j}$ using the method in \cite{liao2019closed}, since the SNS may diversify the numbers of effective paths for different antenna pairs.
Then we introduce the following two ordered sets for the $n$-th DSS
\begin{align} 
\label{equ:OmegaIniI}
  \overline \Omega _n^{\rm{I}} & = \left\{ {q \in \left[ {1,Q} \right]\left| {{T_{{\rm{GI}}}} \le {T_{{\rm{GI}}}} + \frac{{q - 1}}{{{f_{{\rm{ADC}}}}}} < \frac{{{k_n}}}{B}} \right.} \right\}, \\
\label{equ:OmegaIniII}
  \overline \Omega _n^{{\rm{III}}} & = \left\{ {q \in \left[ {1,Q} \right]\left| {\frac{{{k_n}}}{B} + {T_{{\rm{GI}}}}  \le {T_{{\rm{GI}}}} + \frac{{q - 1}}{{{f_{{\rm{ADC}}}}}} < T} \right.} \right\}.
\end{align}
It is easy to verify $\overline \Omega _n^{\rm{I}} \subseteq \Omega _{n}^{\rm{I}} \left(\tau_{n,j,l} \right)$ and $\overline \Omega _n^{\rm{III}} \subseteq \Omega _{n}^{\rm{III}} \left(\tau_{n,j,l} \right)$, $\forall n,j,l$. Therefore both of the sub-matrices $\left[ {{\bf{B}}_{n}} \left({\bm \mu}_{n,j} \right) \right]_{\overline \Omega _n^{{\rm{I}}},:}$ and $\left[ {{\bf{B}}_{n}} \left({\bm \mu}_{n,j} \right) \right]_{\overline \Omega _n^{{\rm{III}}},:}$ will maintain the Vandermonde structure which can be utilized for super-resolution parameter estimation \cite{stoica1989music,liao2019closed}. The set with more elements between $\overline \Omega _n^{\rm{I}}$ and $\overline \Omega _n^{\rm{III}}$ will be chosen as $\Omega _n^{{\rm{ini}}}$ for initial estimation. That is,
\begin{align}\label{equ:SetIni} 
  \Omega _n^{{\rm{ini}}} = \left\{ {\begin{array}{*{20}{l}}
    {\overline \Omega _n^{\rm{I}},}&{{{\left| {\overline \Omega _n^{\rm{I}}} \right|}_{\rm{c}}} \ge {{\left| {\overline \Omega _n^{{\rm{III}}}} \right|}_{\rm{c}}}}\\
    {\overline \Omega _n^{{\rm{III}}},}&{{\rm{otherwise}}}
    \end{array}} \right.,
\end{align}
and the reduced-length measurement ${\left[ {{{\bf{\overline R}}_{n,j}}} \right]_{\Omega _n^{{\rm{ini}}},:}}$ will be treated as the input of the two-dimensional ESPRIT (2D-ESPRIT) algorithm \cite{liao2019closed}. The output of 2D-ESPRIT algorithm, namely, the initial estimates of (probably part of) ${{\bm \mu}_{n,j}}$ and ${{\bm \nu}_{n,j}}$, will be utilized to obtain the initial estimates of (probably part of) ${\bm \alpha}_{n,j}$ as performed in Step 4 of {Algorithm~\ref{ALG1}}.

\subsubsection{Refinement via GDA}

To fully utilize the measurements ${{\bf{\overline R}}_{n,j}}$, we first define an objective function as
\begin{align}\label{equ:OF}  
  g_{n,j}\left({\bm \mu},{\bm \nu},{\bm \alpha}\right) = \left\| { \left( {{\bf{A}}^*\left( {\bm \nu}  \right)}  \odot {{\bf{B}}_n}\left( {\bm \mu}  \right) \right) {\bm \alpha}  - {} {{{\overline {\bf{r}} }_{n,j}}} } \right\|_2^2,
\end{align}
where ${{{\overline {\bf{r}} }_{n,j}}} = {\rm{vec}}\left( {{{\overline {\bf{R}} }_{n,j}}} \right)$.
To find ${\bm \mu}$, ${\bm \nu}$, and ${\bm \alpha}$ that minimize $g_{n,j}\left({\bm \mu},{\bm \nu},{\bm \alpha}\right)$, we resort to the GDA with the initial values obtained by the RL-ESPRIT module, which iteratively refines the current estimates to improve the objective function.
Specifically, in each iteration, the estimates are updated according to Steps 10 and 11 of {Algorithm~\ref{ALG1}}. Due to the page limitation, the expressions of $\frac{{\partial g_{n,j}\left( {{\bm \mu} ,{\bm \nu} ,{\bm \alpha} } \right)}}{{\partial {\bm \mu} }}$ and $\frac{{\partial g_{n,j}\left( {{\bm \mu} ,{\bm \nu} ,{\bm \alpha} } \right)}}{{\partial {\bm \nu} }}$ in Step 8 of {Algorithm~\ref{ALG1}} are omitted. Interested readers are referred to \cite{cui2022channel} for the derivation details. 
{\color{red}
Also note that, to achieve fast convergence in the GDA module, we adopt the Barzilai-Borwein method \cite{barzilai1988two} to compute the step size $\gamma^{(h)}$ in each iteration. Unlike traditional gradient descent with a fixed step size, the Barzilai-Borwein method approximates the behavior of second-order methods with low computational overhead. Specifically, in Step 9 of Algorithm~1, let ${\bm \eta}^{(h)} = {\left[ {{{\left( {{{\bm{\mu }}^{(h)}}} \right)}^{\rm{T}}},{{\left( {{{\bm{\nu }}^{(h)}}} \right)}^{\rm{T}}}} \right]^{\rm{T}}}$ and ${{\bf{g}}^{(h)}} = {\left[ {{{\left( {{\bf{g}}_{{\rm{dis}}}^{(h)}} \right)}^{\rm{T}}},{{\left( {{\bf{g}}_{{\rm{vel}}}^{(h)}} \right)}^{\rm{T}}}} \right]^{\rm{T}}}$, then the Barzilai-Borwein method updates the step size as 
\begin{equation}
\label{equ:BB}
    {\gamma ^{(h)}} = \frac{{{{\left( {{{\bf{g}}^{(h - 1)}} - {{\bf{g}}^{(h - 2)}}} \right)}^{\rm{T}}}\left( {{{\bf{\eta }}^{\left( {h - 1} \right)}} - {{\bf{\eta }}^{\left( {h - 1} \right)}}} \right)}}{{\left\| {{{\bf{g}}^{(h - 1)}} - {{\bf{g}}^{(h - 2)}}} \right\|_2^2}},
\end{equation}
for $h \ge 2$.
}

\subsection{Virtual Bistatic Sensing (VIBS) Paradigm}\label{S4.4}

\begin{figure}[tp!]
\captionsetup{font={footnotesize,color={black}}, name = {Fig.}, singlelinecheck=off, labelsep = period}
\begin{center}
\subfigure[]{\includegraphics[width=0.158\textwidth]{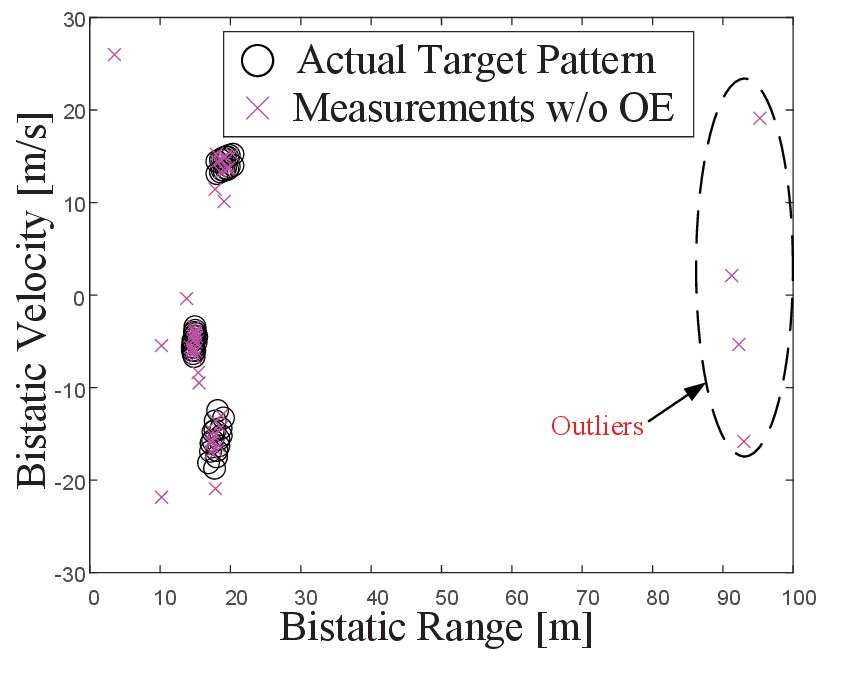}}
\subfigure[]{\includegraphics[width=0.158\textwidth]{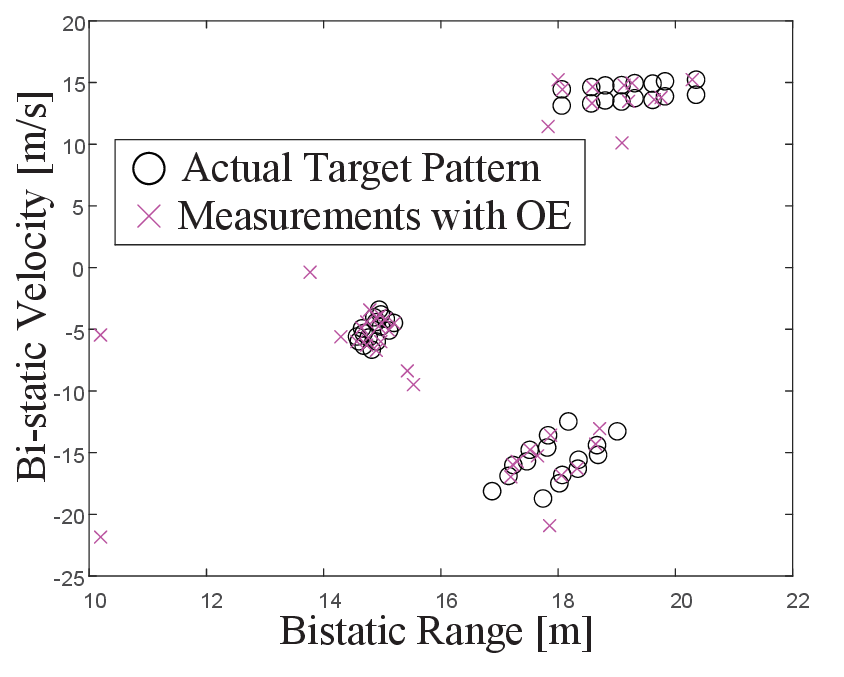}}
\subfigure[]{\includegraphics[width=0.158\textwidth]{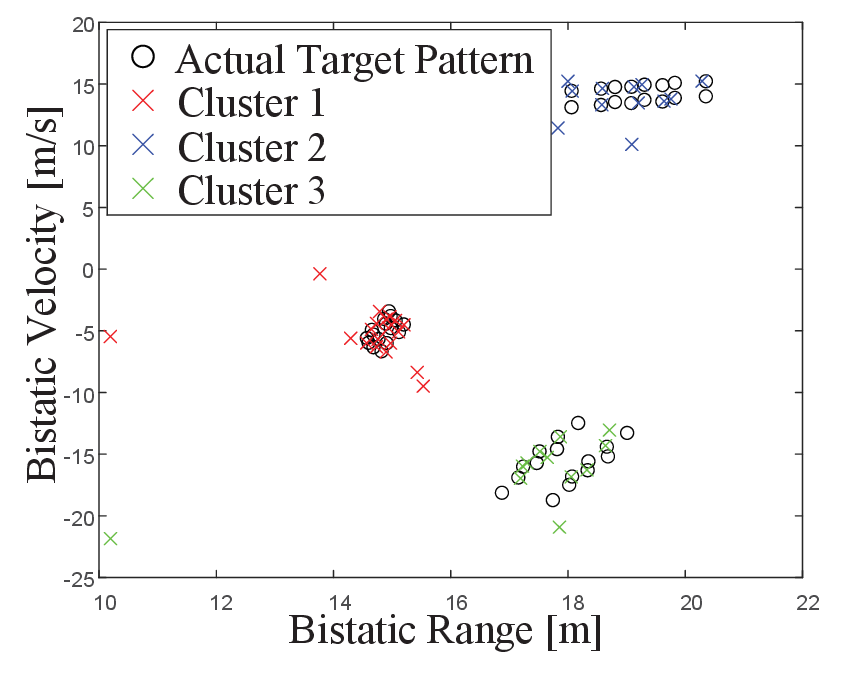}}
\end{center}
\caption{The pre-processing of bistatic sensing with $L = 3$ and $N = N_{\rm Rx} = 4$.
(a) The actual target pattern and the measurements without outlier elimination.
(b) Measurements after outlier elimination.
(c) Results of K-means clustering.
}
\label{fig:PrePro} 
\end{figure}

Based on the results of parameter estimation,  
we can construct the estimates of range and the velocity of the targets associated with the antenna pair $\left(n,j\right)$ as ${\widehat {\bf d}}_{n,j} = \frac{cf_{\rm ADC}T}{B}{\widehat {\bm \mu}}_{n,j}$ and ${\widehat {\bf v}}_{n,j} = \frac{\lambda}{T+T_{\rm GI}+{T_{{\rm{com}}}}}{\widehat {\bm \nu}}_{n,j}$. These estimates can be treated as the measurements from the bistatic radar whose effective transmitter and receiver are the $n$-th DSA and the $j$-th Rx antenna of the sensing receiver, respectively. {\color{red}As discussed in Section~\ref{sec:3}-B, the range and velocity of a near-field target observed by different antenna pairs in ISAC station will be significantly different from each other, even though they are located at a single station. This spatial diversity inspires us to incorporate the virtual bistatic measurements from all antenna pairs to determine the position and velocity for each target, which is called VIBS.}
Before proceeding, we collect ${\widehat {\bf d}}_{n,j}$ and ${\widehat {\bf v}}_{n,j}$ for $n = 1,2,\ldots,N$, $j = 1,2,\ldots,N_{\rm Rx}$ as a data set with
$N_{\rm data} = \sum\nolimits_{n = 1}^N {\sum\nolimits_{j = 1}^{{N_{{\rm{Rx}}}}} {{{\widehat L}_{n,j}}} } $ measurements each corresponding to the estimated bistatic range and velocity of one target. An example of the obtained date set is visualized in Fig.~\ref{fig:PrePro}\,(a). On this basis, we conduct the outlier elimination (OE) to obtain the set of valid measurements, which excludes those measurements that significantly deviate from the normal pattern, as shown in Figs.~\ref{fig:PrePro}\,(a) and \ref{fig:PrePro}\,(b). Then, the clustering algorithm is applied to classify the valid measurements into $\widehat{L}$ clusters, where $\widehat{L} = \mathop {\max }_{n,j} \left\{ {{{\widehat L}_{n,j}}} \right\}$ is the estimate of $L$. For ease of analysis, we let ${\widehat{L}} = L$. Fig.~\ref{fig:PrePro}\,(c) depicts an example of clustering based on K-means method \cite{qian2021k}.
{\color{red}Note that more state-of-the-art clustering algorithms, such as density-based spatial clustering of applications with noise (DBSCAN), can also be adopted to this task.}
For the $l$-th ($l = 1,2,\ldots,L$) cluster, we introduce the following notations:
\begin{itemize}
    \item $P^{\left(l\right)}$: the number of measurements;
    \item ${\widehat d}^{\left(l\right)}_{p}$: the $p$-th ($p = 1,2,\ldots,P^{\left(l\right)}$) range measurement;
    \item ${\widehat v}^{\left(l\right)}_{p}$: the $p$-th ($p = 1,2,\ldots,P^{\left(l\right)}$) velocity measurement;
    \item ${\bf p}^{{\rm Tx},\left(l\right)}_{p} \in \{{\bf p}^{\rm Tx}_{i_1},{\bf p}^{\rm Tx}_{i_2},\ldots,{\bf p}^{\rm Tx}_{i_N}\}$: the position vector of the DSA associated with the $p$-th measurement;
    \item ${\bf p}^{{\rm Rx},\left(l\right)}_{p} \in \{{\bf p}^{\rm Rx}_{1},{\bf p}^{\rm Rx}_{2},\ldots,{\bf p}^{\rm Rx}_{N_{\rm Rx}}\}$: the position vector of the Rx antenna associated with the $p$-th measurement.
\end{itemize}
The measurements can be modeled based on \eqref{equ:DisDefine} and \eqref{equ:VelDefine} as
\begin{align} 
\label{equ:dis}
    {\widehat d^{\left(l\right)}_{p}} & = {\left\| {{{ {\bf{p}} }_l} -  {\bf{p}} _{p}^{{\rm{Tx}},\left(l\right)}} \right\|_2} + {\left\| {{{{\bf{p}} }_l} -  {\bf{p}} _{p}^{{\rm{Rx}},\left(l\right)}} \right\|_2} + e^{{\rm dis},\left(l\right)}_{p},\\
\label{equ:vel}
    {\widehat v^{\left(l\right)}_{p}} & = {\left( {\frac{{{{{\bf{p}} }_l} -  {\bf{p}} _{p}^{{\rm{Tx}},\left(l\right)}}}{{\left\| {{{ {\bf{p}} }_l} -  {\bf{p}} _{p}^{{\rm{Tx}},\left(l\right)}} \right\|_2^{}}} + \frac{{{{{\bf{p}} }_l} -  {\bf{p}} _{p}^{{\rm{Rx}},\left(l\right)}}}{{\left\| {{{{\bf{p}} }_l} -  {\bf{p}} _{p}^{{\rm{Rx}},\left(l\right)}} \right\|_2^{}}}} \right)^{\rm{T}}}{{\bf{v}} _l} + e_{p}^{{\rm{vel}},{\left(l\right)}} ,
\end{align}
where $e_{p}^{{\rm{dis}},{\left(l\right)}}$ and $e_{p}^{{\rm{vel}},{\left(l\right)}}$ are the estimation errors of range and velocity, respectively.
Note that we have assumed that the measurements from the $l$-th cluster corresponds to $l$-th target with position ${\bf p}_l$ and velocity ${\bf v}_l$, which causes no loss of generality. 
The objective of bistatic sensing is to estimate ${\bf p}_l$ and ${\bf v}_l$ based on ${\widehat d_p^{\left( l \right)}}$ and ${\widehat v_p^{\left( l \right)}}$, $p = 1,2,\ldots,P^{\left(l\right)}$. We discuss the estimation of position and velocity separately as follows.

\begin{algorithm}[!t]
\caption{Virtual Bistatic Positioning}
\label{ALG2}
{\bf Input}:
${\widehat d}^{\left(l\right)}_{p}$, ${\widehat v}^{\left(l\right)}_{p}$, ${\bf p}^{{\rm Tx},\left(l\right)}_{p}$, ${\bf p}^{{\rm Rx},\left(l\right)}_{p}$, $p = 1,2,\ldots,P^{\left(l\right)}$ for each $l \in \{1,2,\ldots, L\}$, and maximum number of iterations for GDA $H'_{\max}$.
\begin{algorithmic}[1]
\Statex{\it \% The algorithm will run for each cluster $l \in \{1,2,\ldots, L\}$}
\State For the $l$-th cluster, choose a subset of measurements that involves at least $3$ antenna pairs with the same DSA (or same Rx antenna);
\State Obtain initial position estimate ${\bf \widehat{p}}^{\rm ini}_{l}$ according to Theorem~\ref{Theorem3};
\State $h = 0$, ${{\bf p}}^{\left(0\right)} = {\bf \widehat p}^{\rm ini}_{l}$;
\While{$h < H'_{\max}$,}
    \State $h = h + 1$;
    \State ${\left. {{\bf{g}}_{\rm dis} ^{\left( {h - 1} \right)} = \frac{{\partial {g^{\left(l\right)}}\left( \bf p \right)}}{{\partial {\bf p} }}} \right|_{{\bf p}  = {{\bf p} ^{\left( {h - 1} \right)}}}}$;
    \State {\color{red}Choose step size $\beta^{\left(h\right)}$ based on Barzilai-Borwein method, similarly to Step 9 in Algorithm 1};
    \State ${{\bf p} ^{\left( h \right)}} = {{\bf p} ^{\left( {h - 1} \right)}} - {\beta ^{\left( h \right)}} {\bf g}_{\rm dis}^{\left(h-1\right)}$;
\EndWhile
\end{algorithmic}
{\bf Output}: Refined position estimates ${\bf \widehat{p}}^{\rm GDA}_l = {{\bf p}}^{\left(h\right)}$, $l = 1,2,\ldots,L$.
\end{algorithm}

\subsubsection{Virtual bistatic positioning}

To guarantee a unique target position without ambiguity, the valid measurements should fulfill certain conditions, as stated in Theorem~\ref{Theorem3}.
\begin{theorem}[A sufficient condition for bistatic positioning without ambiguity]
\label{Theorem3}
  For $e^{{\rm dis},\left(l\right)}_{p} = 0$, a sufficient condition for uniquely obtaining ${\bf p}_l$ via $\left\{ {\widehat d_p^{\left( l \right)}} \right\}_{p=1}^{P^{\left(l\right)}}$ is that there exists at least $3$ antenna pairs that involve the same DSA (or same Rx antenna) for the current cluster.
\end{theorem}
\begin{proof}
  See Appendix~\ref{app:c}.
\end{proof}

Theorem~\ref{Theorem3} provides an insight on how to obtain the initial results for VIBS. Specifically, a subset of measurements that involves at least $3$ antenna pairs with the same DSA (or same Rx antenna) can be chosen to obtain the initial (but coarse) estimate of ${\bf \overline p}_l$ according to Theorem~\ref{Theorem3}.
Then, the GDA algorithm can be applied to refine the initial estimate by incorporating all $P^{\left(l\right)}$ measurements. We summarize this virtual bistatic positioning in Algorithm~\ref{ALG2}, where the objective function $g^{\left(l\right)}\left({\bf p}\right)$ in Step 6 is defined by
\begin{align} 
  {g^{\left( l \right)}}\left( {\bf{p}} \right) = \sum\limits_{p = 1}^{{P^{\left( l \right)}}} {\left( {\widehat d_p^{\left( l \right)} - {{\left\| {{\bf{p}} - {\bf{p}}_p^{{\rm{Tx}},\left( l \right)}} \right\|}_2} - {{\left\| {{\bf{p}} - {\bf{p}}_p^{{\rm{Rx}},\left( l \right)}} \right\|}_2}} \right)^2}. 
\end{align}
The gradient of $g^{\left(l\right)}\left({\bf p}\right)$ with respect to $\bf p$ can be readily calculated, and its expression is omitted for brevity.

\subsubsection{Virtual bistatic velocity measurement}

The estimate of velocity vector ${\bf v}_l$ can be easily obtained by replacing ${\bf p}_l$ in \eqref{equ:vel} by its estimate ${\bf \widehat{p}}_l$ obtained in the virtual bistatic positioning. Specifically, when $P^{\left(l\right)}\!\ge\! 3$, the velocity estimate ${\bf \widehat{v}}_l$ is given by
\begin{align}
\label{equ:VE}
    {\bf \widehat{v}}_l\! =\! \left({\bf Y}^{\left(l\right)}\right)^{\dag} {\bf{s}}^{\left(l\right)},
\end{align}
where ${{\bf{s}}^{\left( l \right)}}\! =\! {\left[ {v_1^{\left( l \right)},\ldots,c_{{P^{\left( l \right)}}}^{\left( l \right)}} \right]^{\rm{T}}}\! \in\! \mathbb{R}^{P^{\left(l\right)}}$,
${\bf{Y}}_{}^{\left( l \right)}\! =\! {\left[ {{\bf{y}}_1^{\left( l \right)},\ldots,{\bf{y}}_{{P^{\left( l \right)}}}^{\left( l \right)}} \right]^{\rm{T}}}\! \in\! \mathbb{R}^{P^{\left(l\right)} \times 3}$ and 
${\bf{y}}_p^{\left( l \right)} = \frac{{{{\widehat {\bf{p}}}_l} - {\bf{p}}_p^{{\rm{Tx}},\left( l \right)}}}{{\left\| {{{\widehat {\bf{p}}}_l} - {\bf{p}}_p^{{\rm{Tx}},\left( l \right)}} \right\|_2}} + \frac{{{{\widehat {\bf{p}}}_l} - {\bf{p}}_p^{{\rm{Rx}},\left( l \right)}}}{{\left\| {{{\widehat {\bf{p}}}_l} - {\bf{p}}_p^{{\rm{Rx}},\left( l \right)}} \right\|_2}}$.
Note that unlike the previous works
\cite{wan2024orthogonal, gao2022integrated} which only estimate the radial velocity of the target, the proposed virtual bistatic velocity measurement method can estimate the three-dimensional velocity ${\bf v}_l \in \mathbb{R}^3$, benefiting from the spatial diversity of UM-MIMO and the proposed VIBS paradigm.

{\color{red}
\subsection{Cram\'er-Rao Bound Analysis}
Following the received signal model in \eqref{equ:Rn}, the CRBs for range-dependent parameters ${\bm \mu}_{n,j}$ and velocity-dependent parameters ${\bm \nu}_{n,j}$ are summarized in Theorem~\ref{Theorem4}.

\begin{theorem}[CRB for parameter estimation]
\label{Theorem4}
    Let ${\bm \eta}_{n,j} = \left[ {\bm \mu}_{n,j}^{\rm T}, {\bm \nu}_{n,j}^{\rm T}\right]^{\rm T} \in \mathbb{R}^{2L}$, 
    ${\bf{d}}_n^{{\rm{dis}}}\left( \mu  \right) = \partial {{\bf{b}}_n}\left( \mu  \right)/\partial \mu \in \mathbb{C}^Q$,
    and ${\bf{d}}_{}^{{\rm{vel}}}\left( \nu  \right) = \partial {\bf{a}}\left( \nu  \right)/\partial \nu \in \mathbb{C}^M$. The CRB of ${\bm \eta}_{n,j}$ based on the model \eqref{equ:Rn} can be obtained as
    \begin{align}
        {\rm CRB}\left( {\bm \eta}_{n,j}\right) = \frac{{\sigma _{\rm{n}}^2}}{2}{\left( {{\rm{Re}}\left\{ {{\bf{\Lambda }}_{n,j}^{\rm{H}}{\bf{D}}_{n,j}^{\rm{H}}{\bf{P}}_{n,j}^ \bot {\bf{D}}_{n,j}^{}{\bf{\Lambda }}_{n,j}^{}} \right\}} \right)^{ - 1}},
    \end{align}
    where ${\rm Re}\{ \cdot \}$ extracts the real part of the input, and
    \begin{align*}
        {\bf{\Lambda }}_{n,j}^{} & = {\rm{diag}}\left( {{{\left[ {\bm \alpha _{n,j}^{\rm{T}},\bm \alpha _{n,j}^{\rm{T}}} \right]}^{\rm{T}}}} \right) \in \mathbb{C}^{2L \times 2L}, \nonumber \\
        {\bf{D}}_{n,j}^{} & = \left[ {{{\left( {{\bf{D}}_{n,j}^{\rm vel} } \right)}^*} \odot {{\bf{B}}_n}\left( {{{\bm{\mu }}_{n,j}}} \right),{\bf{A}}_{}^*\left( {{{\bm{\nu }}_{n,j}}} \right) \odot {\bf{D}}_{n,j}^{\rm dis} } \right] \in \mathbb{C}^{MQ \times 2L}, \nonumber \\
        {\bf{P}}_{n,j}^ \bot & = {{\bf{I}}_{MQ}} - {{\bf{C}}_{n,j}} {{\bf{C}}^{\dag}_{n,j}} \in \mathbb{C}^{MQ \times MQ}, \nonumber \\
        {{\bf{C}}_{n,j}} & = { {{\bf{A}}^*\left( {\bm \nu}_{n,j}  \right)} } \odot {{\bf{B}}_n}\left( {\bm \mu}_{n,j}  \right) \in \mathbb{C}^{MQ \times L}, \nonumber \\
        {\bf{D}}_{n,j}^{\rm dis} & = \left[ {{\bf{d}}_{n}^{\rm dis} \left( {{\mu_{n,j,1}}} \right),{\bf{d}}_{n}^{\rm dis} \left( {{\mu_{n,j,2}}} \right),...,{\bf{d}}_{n}^{\rm dis} \left( {{\mu_{n,j,L}}} \right)} \right] \in \mathbb{C}^{Q \times L}, \nonumber \\
        {\bf{D}}_{n,j}^{\rm vel} & = \left[ {{\bf{d}}_{}^{\rm vel} \left( {{\nu_{n,j,1}}} \right),{\bf{d}}_{}^{\rm vel} \left( {{\nu_{n,j,2}}} \right),...,{\bf{d}}_{}^{\rm vel} \left( {{\nu_{n,j,L}}} \right)} \right] \in \mathbb{C}^{M \times L}.
    \end{align*}
\end{theorem}
\begin{proof}
    Based on the vectorized form of \eqref{equ:Rn}, i.e., ${\rm{vec}}\left( {{{\overline {\bf{R}} }_{n,j}}} \right) = {{\bf{C}}_n} \left( {{\bm \mu}_{n,j} ,{\bm \nu}_{n,j} } \right) {\bm \alpha}_{n,j}$, the result can be readily derived according to \cite[Appendix]{vanderveen1998estimation}.
\end{proof}

\subsection{Computational Complexity Analysis}
The computational complexity of the proposed sensing paradigm is analysed in this section. Note that the proposed scheme consists of parameter estimation and VIBS, and their computational complexity in terms of numbers of complex multiplication will be discussed respectively as follows.
\subsubsection{Parameter Estimation (Algorithm~\ref{ALG1})}
The computational complexity of Algorithm~\ref{ALG1} is mainly determined by the following factors.
\begin{itemize}
    \item {\bf 2D-ESPRIT algorithm and complex gain estimation (Steps 3--4)}: It has a computational complexity of approximately ${\cal O}\left( {\max {{\left( {{{\left| {\Omega _n^{{\rm{ini}}}} \right|}_{\rm{c}}},M} \right)}^3 +L^3+MQL}} \right)$ due to the singular value decomposition for ${\left[ {{{\bf{\overline R}}_{n,j}}} \right]_{\Omega _n^{{\rm{ini}}},:}}$ and pseudo-inverse operation.
    \item {\bf GDA refinement (Steps 6--13)}: Each GDA iteration renders complexity about ${\cal O}\left( L^3 + MQL \right)$.
\end{itemize}

\subsubsection{VIBS}
The computational complexity of Algorithm~\ref{ALG2} is mainly determined by its GDA part, which renders complexity about ${\cal O}\left( P^{\left(l\right)} \right)$ for each iteration. Moreover, the 3D velocity estimation in \eqref{equ:VE} also renders complexity about ${\cal O}\left( P^{\left(l\right)} \right)$.
}

%% file: Sec/Sec5_Communications.tex
\section{Sensing-enhanced Near-field CE for UM-MIMO with OCDM}\label{sec:5}

This section focuses on the communication functionality of the ISAC system. Specifically, the near-field CE is formulated as the CS problem under the OCDM waveform, and the CS algorithm is applied to solve it. Then, a sensing-enhanced CE method is proposed, utilizing the obtained sensing results to improve the CE performance.

\subsection{CS-based Channel Estimation}\label{S5.1}

The CE procedure can be implemented during the communication stage as shown in Fig.~\ref{fig:FS}, while preventing the cross-interference between the sensing and communication tasks. Compared to the sensing receiver with low-rate ADCs, the communication receiver at the UT utilizes the Nyquist sampling, and therefore the CE formulation is based on digital representation \cite{ouyang2016orthogonal}. To effectively estimate the finite communication CIR shown in \eqref{CMcom}, we use the shorten OCDM symbols as the pilot signals. Each shorten OCDM symbol consists of $G$ (instead of $K$) orthogonal chirps, where $G \ge \mathop {\max }\limits_l \{\tau_l^{\rm com}\}/T_{\rm s}$.
Moreover, a replica of the shorten OCDM symbol is attached before it to form the cyclic prefix (CP) as the GI, so that each shorten OCDM symbol is of duration $2GT_{\rm s}$.
Based on \eqref{CMcom}, the CIR of interest can be obtained as $\left\{ {{\bf{h}}_g^{{\rm{com}}} = {{\bf{h}}^{{\rm{com}}}}\left( {g {T_{\rm{s}}}} \right)} \right\}_{g = 0}^{G-1}$. Note that the length of GI (CP) guarantees that ${{\bf{h}}_g^{{\rm{com}}}} = {\bf 0}$ when $g \ge G$. 
We denote the number of pilot symbols for CE as $N_{\rm CE}$, and the $p$-th transmit signal as
\begin{align} 
  {\bf S}_p = {\bf \Phi}^{\rm H}{\bf X}_p,
\end{align}
where $p = 1,2,\ldots, N_{\rm CE}$, ${\bf \Phi} \in \mathbb{C}^{G \times G}$ is the $G$-order discrete Fresnel transform (DFnT) matrix with the elements ${\left[ {{{\bf{\Phi }}}} \right]_{i,j}} = e^{ - \textsf{j}\frac{\pi }{4}} {{e^{\textsf{j}\frac{\pi }{G}{{\left( {i - j} \right)}^2}}}} / {\sqrt G }$ \cite{ouyang2016orthogonal}, and ${\bf X}_p \in \mathbb{C}^{G \times N_{\rm Tx}}$ is the normalized transmitted constellation symbols.  ${\bf S}_p$ can be further denoted as ${{\bf{S}}_p} = {\left[ {{\bf{s}}_{p,0}^{},{\bf{s}}_{p,1}^{},\ldots,{\bf{s}}_{p,G-1}^{}} \right]^{\rm{T}}}$, where ${\bf{s}}_{p,g} \in \mathbb{C}^{N_{\rm Tx}}$ are the modulated signals for all antennas. Given the application of CP, the received pilot signals of $p$-th pilot signal at the $g$-th sample, $y_{p,g}$, can be formulated based on circular convolution between the CIR and modulated signals. That is,
\begin{align} 
  {y_{p,g}} = \sqrt{P_{\rm Tx}^{\rm com}}\sum\limits_{g' = 0}^{G - 1} {{\bf{s}}_{p,{{\left\langle {g - g'} \right\rangle }_G}}^{\rm{T}}{\bf{h}}_{g'}^{{\rm{com}}}} + n_{p,g},
\end{align}
where $P_{\rm Tx}^{\rm com}$ is the transmit power of ISAC station during the communication stage, $n_{p,g} \sim {\cal CN}\left(0,\sigma_{\rm n}^2\right)$ is the AWGN.
Stacking the $G$ samples of received signals yields
\begin{align}\label{equ:RxCom} 
  {\bf y}_{p}\! & =\! {\left[ {{y_{p,0}},{y_{p,1}},\ldots,{y_{p,G - 1}}} \right]^{\rm{T}}} 
  \! =\! \sqrt{P_{\rm Tx}^{\rm com}}{{\bf{C}}_p}{{\bf{h}}_{{\rm{CIR}}}}\! +\! {\bf n}_p,\!
\end{align}
where ${\bf n}_p = {\left[ {{n_{p,0}},{n_{p,1}},\ldots,{n_{p,G - 1}}} \right]^{\rm{T}}} \in \mathbb{C}^{G}$, ${{\bf{h}}_{{\rm{CIR}}}} = {\left[ {{{\left( {{\bf{h}}_0^{{\rm{com}}}} \right)}^{\rm{T}}},{{\left( {{\bf{h}}_1^{{\rm{com}}}} \right)}^{\rm{T}}},\ldots,{{\left( {{\bf{h}}_{G - 1}^{{\rm{com}}}} \right)}^{\rm{T}}}} \right]^{\rm{T}}} \in \mathbb{C}^{GN_{\rm Tx}}$ is the overall CIR matrix to be estimated, and
\begin{align} 
  {{\bf{C}}_p} = \left[ {\begin{array}{*{20}{c}}
{{\bf{s}}_{p,0}^{\rm{T}}}&{{\bf{s}}_{p,G - 1}^{\rm{T}}}& \cdots &{{\bf{s}}_{p,1}^{\rm{T}}}\\
{{\bf{s}}_{p,1}^{\rm{T}}}&{{\bf{s}}_{p,0}^{\rm{T}}}& \cdots &{{\bf{s}}_{p,2}^{\rm{T}}}\\
 \vdots & \vdots & \ddots & \vdots \\
{{\bf{s}}_{p,G - 1}^{\rm{T}}}&{{\bf{s}}_{p,G - 2}^{\rm{T}}}& \cdots &{{\bf{s}}_{p,0}^{\rm{T}}}
\end{array}} \right] \in \mathbb{C}^{G \times GN_{\rm Tx}}.
\end{align}
Note that ${\bf C}_p$ is a block-circulant matrix \cite{olson2014circulant}. To facilitate CE, we transform the CIR vector $h_{\rm CIR}$ into the frequency domain (marked by ``Fd'') by applying discrete Fourier transform (DFT)
\begin{align} 
  {{\bf{h}}^{{\rm{Fd}}}_g} = \frac{1}{\sqrt{G}} \sum\nolimits_{g' = 0}^{G - 1} {{\bf{h}}_{g'}^{{\rm{com}}}{e^{-\textsf{j}\frac{{2\pi }}{G}gg'}}}.
\end{align}
This can be written more compactly as
\begin{align}\label{equ:hFd} 
  {{\bf{h}}_{{\rm{Fd}}}} & = {\left[ {{{\left( {{\bf{h}}_0^{{\rm{Fd}}}} \right)}^{\rm{T}}},{{\left( {{\bf{h}}_1^{{\rm{Fd}}}} \right)}^{\rm{T}}},...,{{\left( {{\bf{h}}_{G - 1}^{{\rm{Fd}}}} \right)}^{\rm{T}}}} \right]^{\rm{T}}} \nonumber \\
    &= \left({\bf F}_{\rm DFT} \otimes {\bf I}_{N_{\rm Tx}} \right) {{\bf{h}}_{{\rm{CIR}}}},
\end{align}
where ${\bf F}_{\rm DFT}$ is the $G$-order normalized DFT matrix.
Estimating the CIR can be equivalently recast into estimating ${\bf h}_{\rm Fd}$. Based on \eqref{equ:hFd}, we re-write \eqref{equ:RxCom} as
\begin{align} 
  {{\bf{y}}_p} = \sqrt {P_{{\rm{Tx}}}^{{\rm{com}}}} {{\bf{C}}_p}\left( {{\bf{F}}_{{\rm{DFT}}}^{\rm{H}} \otimes {{\bf{I}}_{{N_{{\rm{Tx}}}}}}} \right){{\bf{h}}_{{\rm{Fd}}}} + {\bf{n}}_p.
\end{align}
To fully utilize the block-circulant property of ${\bf C}_p$ and thus simplify the CE problem, the UT conducts the DFT for the received signals ${\bf y}_p$ and obtain
\begin{align}\label{equ:DFTcom} 
  {{{\bf{\overline y}}}_p} = {{\bf{F}}_{{\rm{DFT}}}}{{\bf{y}}_p} = \sqrt {P_{{\rm{Tx}}}^{{\rm{com}}}} {{\bf{\Lambda }}_p} {{\bf{h}}_{{\rm{Fd}}}} + {\bf{\overline n}}_p,
\end{align}
where ${\bf \overline{n}}_p = {\bf{F}}_{{\rm{DFT}}}{\bf{n}}_p$, and ${{\bf{\Lambda }}_p} = {{{\bf{F}}_{{\rm{DFT}}}}{{\bf{C}}_p}\left( {{\bf{F}}_{{\rm{DFT}}}^{\rm{H}} \otimes {{\bf{I}}_{{N_{{\rm{Tx}}}}}}} \right)}$. According to the property of block-circulant matrices \cite{olson2014circulant}, ${{\bf{\Lambda }}_p} \in \mathbb{C}^{G \times GN_{\rm Tx}}$ is a block-diagonal matrix, which can be represented by
\begin{align} 
  {{\bf{\Lambda }}_p} = {\rm blkdiag}\left( {{\bf{d}}_{p,0}^{\rm{T}},{\bf{d}}_{p,1}^{\rm{T}},\ldots,{\bf{d}}_{p,G - 1}^{\rm{T}}} \right),
\end{align}
with ${\bf{d}}_{p,g}^{} = \sum\nolimits_{g' = 0}^{G - 1} {{\bf{s}}_{p,g'}^{}{e^{ - \textsf{j}\frac{{2\pi }}{G}gg'}}} $. In this way, $G$ frequency-domain channels are sounded independently. Specifically, \eqref{equ:DFTcom} can be written as 
\begin{align} 
  {\overline y}_{p,g} = \sqrt{P^{\rm com}_{\rm Tx}} {\bf d}^{\rm T}_{p,g}{\bf h}_g^{\rm Fd} + {\overline n}_{p,g},
\end{align}
where ${\overline y}_{p,g}$ (${\overline n}_{p,g}$) is the $\left(g+1\right)$-th element of ${\bf \overline y}_{p}$ (${\bf{\overline n}}_p$). By collecting $N_{\rm CE}$ pilot signals, the final observations for each frequency-domain channel can be obtained as
\begin{align}\label{equ:CSpro1} 
  {\bf y}^{\rm com}_g = \sqrt{P^{\rm com}_{\rm Tx}} \mathbf{D}_{g} \mathbf{h}_{g}^{\rm Fd} + {{\mathbf{n}}_{g}^{\rm com}},
\end{align}
where ${\bf{y}}_g^{{\rm{com}}}\! =\! {\left[ {{{\overline y}_{1,g}},{{\overline y}_{2,g}},\ldots ,{{\overline y}_{{N_{{\rm{CE}}}},g}}} \right]^{\rm{T}}}\! \in\! \mathbb{C}^{N_{\rm CE}}$, ${\mathbf{D}_{g}\! =\! {\left[ {{\bf{d}}_{1,g}^{},{\bf{d}}_{2,g}^{},\ldots,{\bf{d}}_{{N_{{\rm{CE}}}},g}^{}} \right]^{\rm{T}}}\! \in\! \mathbb{C}^{N_{\rm CE}\times N_{\rm Tx}}}$, and ${\mathbf{n}^{\rm com}_{g}} =$ ${\left[ {\overline n_{1,g},\overline n_{2,g},\ldots,\overline n_{{N_{{\rm{CE}}}},g}} \right]^{\rm{T}}}\! \in\! \mathbb{C}^{N_{\rm CE}}$.

\subsection{Initial Compressive Sensing Based CE}\label{S5.2}

It can be observed that when the number of pilot signals $N_{\rm CE}$ is small, obtaining the estimate of $\mathbf{h}_{g}^{\rm Fd}$ from \eqref{equ:CSpro1} becomes underdetermined, since the dimension of $\mathbf{h}_{g}^{\rm Fd}$ is typically much larger than that of observations ${\bf y}^{\rm com}_g$, i.e., $N_{\rm CE} \ll N_{\rm Tx}$. In this case, conventional channel estimation methods, such as least square (LS) and minimum mean square error (MMSE), cannot provide reliable solutions due to the lack of sufficient pilot observations.
Fortunately, the inherent sparsity of near-field channel enables reformulating the CE problem based on the framework of CS, where sparse recovery algorithms can be exploited to accurately estimate $\mathbf{h}_{g}^{\rm Fd}$ even with limited pilot overhead. 
To this end, we introduce a two-dimensional polar-domain transform matrix $\mathbf{W}\in \mathbb{C}^{N_{\rm Tx} \times N_{\rm Pd}}$, whose columns are constructed by $N_{\rm Pd} \ge N_{\rm Tx}$ near-field steering vectors parameterized by both the distance and the virtual angle. In this way, $\mathbf{W}$ captures the two-dimensional geometry of the near-field channel.
Due to the page limitation, the specific formulation of $\mathbf{W}$ is omitted here, and interested readers are referred to \cite[Algorithm 1]{cui2022channel}.
By applying this transform, each frequency-domain near-field channel can be represented as
\begin{align}\label{equ:FdCh} 
  \mathbf{h}^{\rm Fd}_{g} = \mathbf{W}\, \mathbf{h}_{g}^{\rm Pd},
\end{align}
where $\mathbf{h}_{g}^{\rm Pd}\in \mathbb{C}^{N_{\rm Pd}}$ denotes the polar-domain channel vector with sparsity. Substituting \eqref{equ:FdCh} into \eqref{equ:CSpro1} yields
\begin{align}\label{CSpro2} 
  {\bf y}^{\rm com}_g = \sqrt{P^{\rm com}_{\rm Tx}} \mathbf{D}_{g} \mathbf{W} \mathbf{h}_{g}^{\rm Pd} + {\overline{\mathbf{n}}_{g}^{\rm com}},
\end{align}
for $g = 0,1,\ldots,G-1$, which gives a series of canonical CS problems. The $G$ polar-domain channel vectors $\{\mathbf{h}_{g}^{\rm Pd}\}_{g=0}^{G-1}$ sharing the same sparsity pattern are observed via $G$ different sensing matrices $\{\mathbf{D}_{g} \mathbf{W}\}_{g=0}^{G-1}$, which renders the generalized multiple measurement vector (GMMV) problem \cite{gao2015spatially}. Therefore, it can be effectively solved by off-the-shelf CS algorithms that jointly leverage the common sparsity pattern. We utilize the DOMP algorithm \cite{wan2020broadband} to estimate the $G$ polar-domain channel vectors, which is summarized in Steps 1--9 of Algorithm~\ref{ALG3}.

\subsection{CE Refinement Based on Sensing Results}\label{S5.3}

\begin{algorithm}[!t]
\caption{Near-field CE and Its Enhancement Based on Sensing Results}
\label{ALG3}
\begin{algorithmic}[1]
\renewcommand{\algorithmicrequire}{\textbf{Input:}}
\renewcommand{\algorithmicensure}{\textbf{Output:}}
\Require {Measurements $\{{{\bf{y}}_g^{\rm com}}\}_{g=0}^{G-1}$, pilot-determined matrices $\{{\bf D}_g\}_{g=0}^{G-1}$, polar-domain transform matrix ${\bf W}$, sensing results $\{{\bf \widehat p}_l\}_{l=1}^{L}$, and pre-defined threshold $\varepsilon$.}
\Statex {\it \% Initial CE based on DOMP begins}
\State Initialization: $\mathcal{I} = \emptyset$, ${{\bf{r}}_g} = {{\bf{y}}^{\rm com}_g}$, and ${\bf{\widehat h}}_{g}^{\rm Pd} = {\bf{0}}_{N_{\rm Pd} \times 1}$;
\While{{\it Stopping Criterion} is not met}
    \State ${i^*} = \arg \max\limits_i \sum\nolimits_{g = 0}^{G - 1} {\left| {{{\left[ {{{\left( {{{\bf{D}}_g}{\bf{W}}} \right)}^{\rm H}}{\bf{r}}_g} \right]}_i}} \right|} $;
    \State ${{\mathcal{I}}}={{\mathcal{I}}} \cup \{{{i}^{*}}\}$;
    \State ${\bf{\widehat h}}_{g} = \left[ {{{\bf{D}}_{g}}{\bf{W }}} \right]_{\cal I}^\dag {{\bf{y}}_g^{\rm com}}$ for $g = 0,1,\ldots,G-1$;
    \State ${{\bf{r}}_g} = {{\bf{y}}_g} - \left[ {{\bf{D}}_{g}}{\bf{W}}\right]_{\cal I}{{\bf{\widehat h}}^{\rm Pd}_g}$ for $g = 0,1,\ldots,G-1$;
\EndWhile
\State ${\left[ {{{{\bf{\widehat h}}}^{\rm Pd}_{g}}} \right]_{\cal I}} = {{{\bf{\widehat h}}}_g}$ for $g = 0,1,\ldots,G-1$;
\State ${\bf \widehat h}_g^{\rm Fd,ini} = {\bf W}{{{{\bf{\widehat h}}}^{\rm Pd}_{g}}}$ for $g = 0,1,\ldots,G-1$;
\Statex {\% \it Initial CE ends and the enhancement begins}
\State Initialization: ${\cal J} = \emptyset$, ${\bf W}_{\rm sel} = \left[{\bf W}\right]_{\cal I}$;
\For{$l = 1,2,\ldots,L$}
\State Generate vector ${{{\bf a}_{\rm sen}}}\in \mathbb{C}^{N_{\rm Tx}}$ with $\left[{{{\bf a}_{\rm sen}}}\right]_i = e^{-\textsf{j}{2\pi}{f_{\rm c}}(\left\| {{\bf \widehat p}_l - {\bf{p}}_i^{{\rm{Tx}}}} \right\|_2^{} - \left\| {{\bf \widehat p}_l} \right\|_2)}/\sqrt{N_{\rm Tx}}$, $i=1,2,\ldots,N_{\rm Tx}$;
\State ${\cal J} = {\cal J} \cup \left\{ {j\left| {{{\left[ {{\bf{W}}_{\rm sel}^{\rm{H}}{{\bf{a}}_{{\rm{sen}}}}} \right]}_j} \ge \varepsilon } \right.} \right\}$;
\State $\left[{\bf W}_{\rm sel}\right]_{\cal J} = {{{\bf a}_{\rm sen}}}$;
\EndFor
\State ${\bf \widehat h}^{\rm coeff}_g = \left({\bf D}_g {\bf W}_{\rm sel}\right)^\dag{\bf y}^{\rm com}_g$ for $g = 0,1,\ldots,G-1$;
\State ${\bf \widehat h}^{\rm Fd,enh}_g = {\bf W}_{\rm sel}{\bf \widehat h}^{\rm coeff}_g$ for $g = 0,1,\ldots,G-1$;
\Ensure {Initial CE results ${\bf \widehat h}_g^{\rm Fd,ini}$ and their improved version ${\bf \widehat h}^{\rm Fd,enh}_g$, $g = 0,1,\ldots,G-1$.}
\end{algorithmic}
\end{algorithm}

The initial channel estimates ${\bf \widehat h}_g^{\rm Fd}$ obtained from Algorithm~\ref{ALG3} can be used for beamforming and equalization in communication stages (see Fig. \ref{fig:FS}). However, the on-grid DOMP suffers from limited spatial resolution, which may make communication performance unsatisfactory.
Fortunately, given that the communication channel and sensing channel may share some common scatterers \cite{liu2020joint}, the sensing results obtained from the sensing process (as detailed in the previous sections) can help enhance the CE.
The enhancement process is summarized in Steps 10--17 of Algorithm~\ref{ALG3}. It consists of mainly two following stages.
\begin{itemize}
    \item {\it Sensing results matching and replacement (Steps 12-14)}. For each position coordinate obtained from the sensing process, the near-field steering vector is generated according to \eqref{NFSV}, and then compared to all the selected codewords (i.e., ${\bf W}_{\rm sel}$ in Step 10) in the polar-domain dictionary via correlation operation. A threshold $\varepsilon$ is set to evaluate the similarity, and the selected codewords that are similar to the sensing-associated near-field steering vector are recorded. Then, these codewords are replaced by the near-field steering vectors from sensing results to obtain the refined support (Step 14), which are more precise to represent the communication channels.
    \item {CE enhancement based on refined support (Step 16)}. With the refined support, the observations in \eqref{CSpro2} become an overdetermined system, so that the polar-domain channel coefficients can be updated via LS, as done in Step 16 of Algorithm~\ref{ALG3}. The estimate of enhanced frequency-domain channels $\{{\bf \widehat h}^{\rm Fd,enh}_g\}_{g=0}^{G-1}$ can be obtained by multiplying the refined support matrix by the updated coefficients, as done in Step 17 of Algorithm~\ref{ALG3}.
\end{itemize}

{\color{red}
\begin{remark}
To provide a holistic view of the proposed architecture, it is essential to highlight how the communication and sensing functionalities seamlessly coexist and cooperate. For coexistence, communication OCDM symbols fill the gaps between sensing symbols to maximize time-domain utilization (Fig. 3). For cooperation, the dual functions are tightly linked via the proposed sensing-enhanced CE scheme. By providing extracted near-field physical parameters (i.e., positions of scatterers) as crucial spatial priors, the proposed CE scheme refines the polar-domain channel support set and substantially facilitates CE for UM-MIMO communications.
\end{remark}

Also note that the initial CE in Algorithm~\ref{ALG3} is based on state-of-the-art OMP-type algorithms, so its computational complexity can be found in literature like \cite{gao2022integrated,wan2020compressive}. Moreover, the sensing enhancement procedure in Algorithm~\ref{ALG3} renders complexity ${\cal O}\left( {N_{{\rm{Tx}}}}L + \left| \cal{I} \right|_{\rm{c}}^3 + {N_{{\rm{CE}}}}\left| \cal{I}  \right|_{\rm{c}}^2 + {N_{{\rm{Tx}}}}\left| \cal{I}  \right|_{\rm{c}}^{} \right)$ at most, since in general not all targets detected by the sensing phase play the role of scatterers for communications.
}

%% file: Sec/Sec6_Simulation.tex
\section{Simulation Results}\label{sec:6}

In this section, we present simulation results to evaluate the performance of the proposed ISAC scheme, and compare it with existing counterparts in the literature.

\subsection{Experimental Setting}\label{S6.1}

The default simulation system parameters are listed in Table~\ref{TAB1}. The signal-to-noise ratio (SNR) is defined by $P_{\rm Tx}/\sigma_{\rm n}^2$ for sensing and by $P_{\rm Tx}^{\rm com}/\sigma_{\rm n}^2$ for communications, where $\sigma_{\rm n}^2$ is the power of AWGN. 
According to Table~\ref{TAB1} and \eqref{equ:solution}, the DSS selection must satisfy $48 \le \left| {{k_{n'}} - {k_n}} \right| \le 208$, $\forall n \neq n'$. Therefore, we construct ${\cal I}_{\rm DSS}$ by randomly selecting $N$ elements in $\left\{ {0,48,96,144,192} \right\}$.
The model of sensing channel complex gains $\alpha_{i,j,l}$ differs based on the considered channel environments as discussed in Section~\ref{S3.2}. Specifically,
\begin{itemize}
    \item For the spatially-correlated channels, $\alpha_{i,j,l} = \alpha_{l} \sim {\cal CN}\left(0,1\right)$, $\forall i,j$.
    \item For the SUC channels, $\alpha_{i,j,l} \sim {\cal CN} \left( 0, 1 \right)$, and we adopt this channel model unless stated otherwise.
    \item When SNS is considered, we randomly choose $N_{\rm null}$ out of $NL$ Tx-target paths in the SUC channels and make the corresponding complex gains zero. 
\end{itemize}
The communication channel is generated in the same way as the sensing channel.
The polar-domain dictionary $\bf W$ is generated according to \cite[Algorithm 1]{cui2022channel} within the distance $\left[3 ,20\right]$\,m and the oversampling rate for angle quantization is set to $2$.
Algorithm~\ref{ALG3} is terminated when the number of iterations reaches $10$.
The K-means method is adopted as the default clustering algorithm for VIBS.
Other settings and the benchmark schemes will be specified during the presentation of results.

\begin{table}[t]
\centering
\captionsetup{font={footnotesize,color={black}},labelsep=newline}
\caption{\sc Default System Parameters}
\label{TAB1}
\vspace*{-1mm}
\resizebox{\columnwidth}{!}{
\begin{tabular}{l l l}
\hline
\textbf{Symbol} & \textbf{Description} & \textbf{Value} \\
\hline
$f_{\rm c}$ & Carrier frequency & $30$\,GHz \\
$\lambda$ & Wavelength & $0.01$\,m \\
$K$ & Number of OCDM subcarriers & $256$ \\
$B$ & Bandwidth & $100$\,MHz \\
$M$ & Number of sensing symbols & $64$ \\
$T_{\rm GI}$ & Guard interval & $0.16\,\mu$s \\
$f_{\rm LPF}=f_{\rm ADC}$ & LPF cutoff frequency/ADC sampling rate & $12.5$\,MHz \\
$P_{\rm Tx} = P_{\rm Tx}^{\rm com}$ & Transmit power for sensing/communication & $50$\,dBm \\
$N_{\rm Tx}$ & Number of Tx antennas & $512$ \\
${\bf p}^{\rm Tx}_i$ & Coordinate of the $i$-th Tx antenna & $\big[(i\!-\!1)\lambda/2,0,0\big]^{\rm T}$ \\
$N_{\rm Rx}$ & Number of Rx antennas & $4$ \\
$N$ & Number of DSSs/DSAs & $4$ \\
$D_{\rm Rx}$ & Distance between adjacent Rx antennas & $1$\,m \\
${\bf p}^{\rm Rx}_j$ & Coordinate of Rx antennas & $[\pm D_{\rm Rx}/2,\pm D_{\rm Rx}/2,0]^{\rm T}$ \\
$L$ & Number of sensing targets & $3$ \\
$r_l$ & Distance & ${\cal U}(5,10)$\,[m] \\
$\theta_l$ & Elevation angle & ${\cal U}(0,\pi/2)$\,[rad] \\
$\phi_l$ & Azimuth angle & ${\cal U}(0,2\pi)$\,[rad] \\
$v_{x,l},v_{y,l},v_{z,l}$ & Velocities & ${\cal U}(0,100)$\,[km/h] \\
{\color{red}$H_{\max}$} & {\color{red}Maximum iterations in Algorithm~\ref{ALG1}} & {\color{red}$30$} \\
{\color{red}$H'_{\max}$} & {\color{red}Maximum iterations in Algorithm~\ref{ALG2}} & {\color{red}$10$} \\
$T_{\rm com}$ & Duration of communication stage & $10.9\,\mu$s \\
$G$ & CP length & $32$ \\
$\varepsilon$ & Threshold in Algorithm~\ref{ALG3} & $0.6$ \\
\hline
\end{tabular}
}
\vspace*{-3mm}
\end{table}

\subsection{Numerical Results}\label{S6.2}

\begin{figure}[bp!]
\captionsetup{font={footnotesize}, name = {Fig.}, singlelinecheck=off, labelsep = period}
\begin{center}
\subfigure[]{\includegraphics[width=0.24\textwidth]{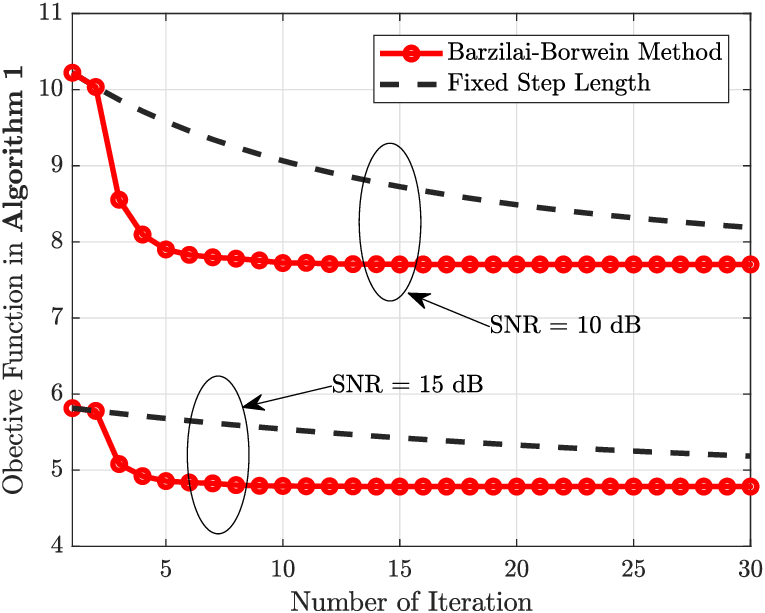}}
\subfigure[]{\includegraphics[width=0.24\textwidth]{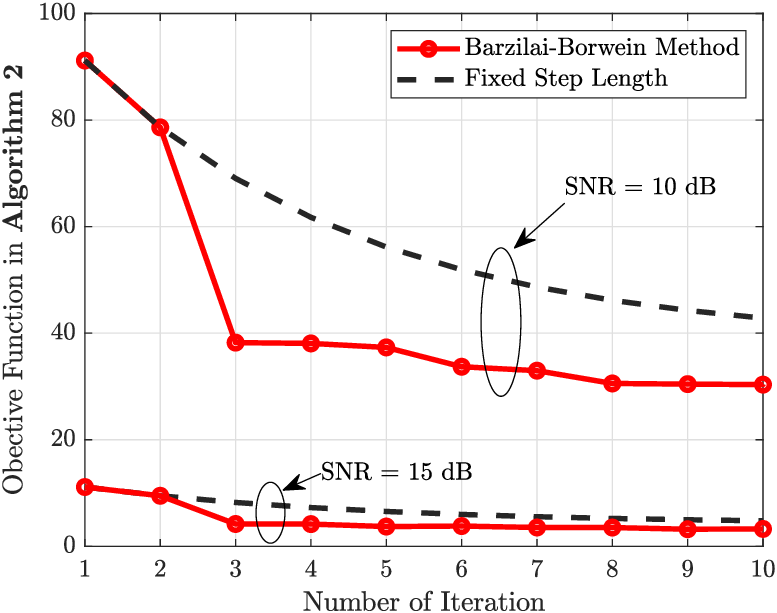}}
\end{center}
\caption{The convergence behaviour of (a) parameter estimation in Algorithm~\ref{ALG1}, and (b) virtual bistatic positioning Algorithm~\ref{ALG2}. 
}
\label{fig:cvg} 
\end{figure}

\subsubsection{Sensing performance}

First, we investigate the convergence performance of the proposed parameter estimation method and the virtual bistatic positioning, as illustrated in Fig.~\ref{fig:cvg}. It is evident that the objective functions in both Algorithm~\ref{ALG1} and Algorithm~\ref{ALG2} decrease over iteration, and the convergence occurs much more rapidly when employing the Barzilai-Borwein method compared to the conventional method with a fixed step size. {\color{red}Based on the results in Fig.~\ref{fig:cvg}, we adopt a fixed maximum number of iterations as shown in Table~\ref{TAB1} for the following simulation to ensure deterministic execution time for real-time hardware implementations.}

\begin{figure}[tp!]
\captionsetup{font={footnotesize,color={black}}, name = {Fig.}, singlelinecheck=off, labelsep = period}
\begin{center}
\subfigure[]{\includegraphics[width=0.25\textwidth]{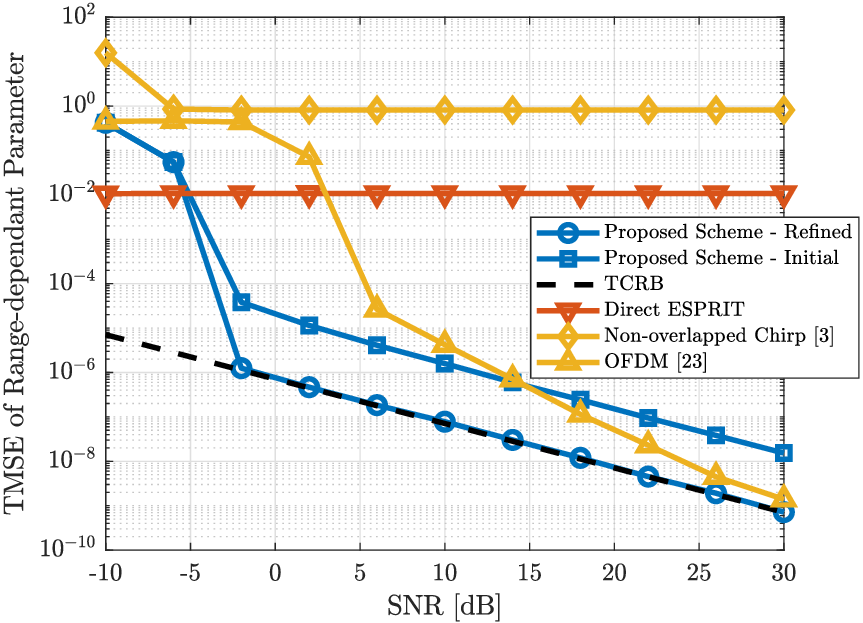}}
\subfigure[]{\includegraphics[width=0.23\textwidth]{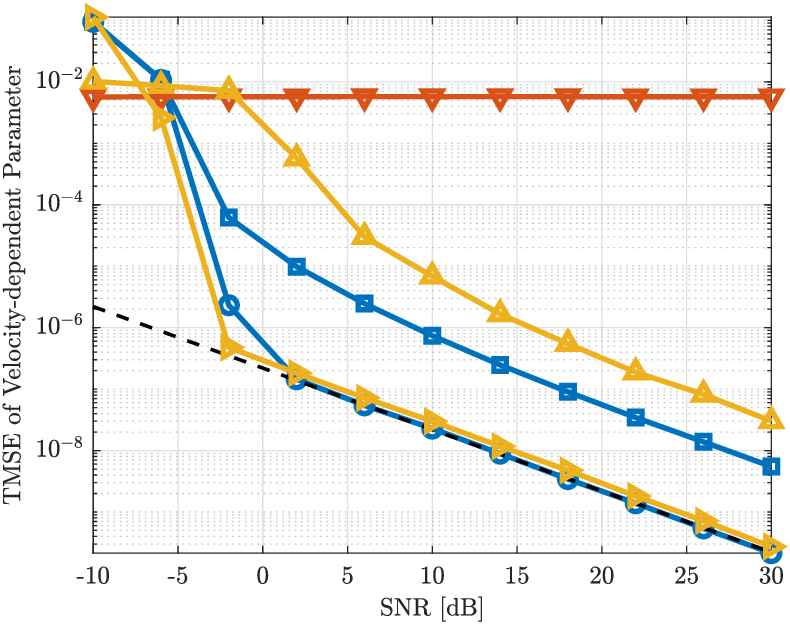}}
\end{center}
\vspace{-4mm}
\caption{The TMSE performance of (a) range-dependent parameter estimation and (b) velocity-dependent parameter estimation. The legend is shared by both figures.
}
\label{fig:MSE_tf} 
\vspace*{-4mm}
\end{figure}

Then, we investigate the total mean square error (TMSE) performance of parameter estimation. The TMSEs of range-dependent and velocity-dependent parameter estimations are defined as ${\mathbb E}\left\{\sum\nolimits_{n = 1}^N {\sum\nolimits_{j = 1}^{{N_{{\rm{Rx}}}}} {\left\| {{{\widehat {\bm{\mu }}}_{n,j}} - {{\bm{\mu }}_{n,j}}} \right\|_2^2} }\right\}$ and ${\mathbb E}\left\{\sum\nolimits_{n = 1}^N {\sum\nolimits_{j = 1}^{{N_{{\rm{Rx}}}}} {\left\| {{{\widehat {\bm{\nu }}}_{n,j}} - {{\bm{\nu }}_{n,j}}} \right\|_2^2} }\right\}$, respectively, where ${{\widehat {\bm{\mu }}}_{n,j}}$ (${{\widehat {\bm{\nu }}}_{n,j}}$) is the estimate of ${{{\bm{\mu }}}_{n,j}}$ (${{{\bm{\nu }}}_{n,j}}$). To validate the effectiveness of the estimate, we also introduce the total CRBs (TCRBs) defined as $\sum\nolimits_{n = 1}^N {\sum\nolimits_{j = 1}^{{N_{{\rm{Rx}}}}} {{\rm{CRB}}\left( {{\bm \mu _{n,j}}} \right)} }$ and $\sum\nolimits_{n = 1}^N {\sum\nolimits_{j = 1}^{{N_{{\rm{Rx}}}}} {{\rm{CRB}}\left( {{\bm \nu _{n,j}}} \right)}}$, where the specific CRB expression can be obtained by applying the results in \cite[Appendix]{vanderveen1998estimation} to the signal model \eqref{equ:Rn}.
Figs.~\ref{fig:MSE_tf}\,(a) and (b) show the TMSEs for range-dependent and velocity-dependent parameter estimations, respectively, as the functions of SNR. Notably, due to the dual discontinuity in the OCDM-FMCW processing, directly applying ESPRIT to the digitized IF signal (marked by `Direct ESPRIT') is ineffective. In contrast, the proposed two-stage parameter estimation method performs well despite the dual discontinuity, with GDA-based refinement enhancing estimation accuracy and enabling TMSE to closely approach the TCRB.
We also consider the following two state-of-the-art waveform benchmarks in Fig.~\ref{fig:MSE_tf}: (i) The OFDM waveform \cite{sturm2011waveform} with $K$ subcarriers and bandwidth $B$; and (ii) the non-overlapped chirp waveform design \cite{liu2020joint}, which divides the whole bandwidth into $N$ non-overlapped segments to transmit $N$ orthogonal chirp signals (see \cite[Fig. 4]{liu2020joint}).
For these two benchmarks, we utilize 2D-ESPRIT for parameter extraction.
{\color{red}Although the OFDM waveform yields much more range-related measurements (of length $K = 256$) than the proposed scheme (of length $Q = 30$), the latter outperforms the former with reduced hardware complexity.
Also note that required ADC sampling rate for the proposed scheme ($12.5$\,MHz) is much lower than that for traditional schemes ($100$\,MHz), which indicates about eightfold reduction for ADC power consumption.}
Moreover, the non-overlapped chirp method \cite{liu2020joint} exhibits poor range-dependent parameter estimation performance due to the range resolution degradation.

\begin{figure}[tp!]
\captionsetup{font={footnotesize,color={red}}, name = {Fig.}, singlelinecheck=off, labelsep = period}
\begin{center}
\subfigure[]{\includegraphics[width=0.241\textwidth]{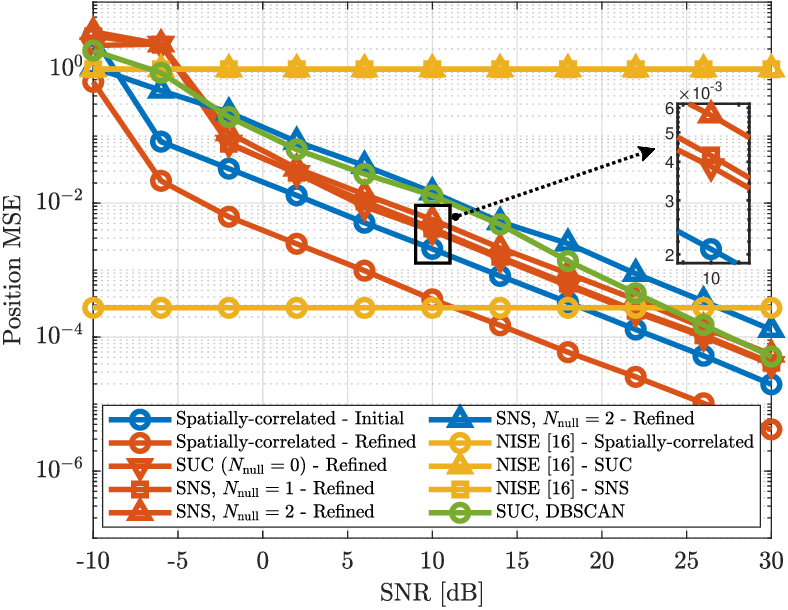}}
\subfigure[]{\includegraphics[width=0.241\textwidth]{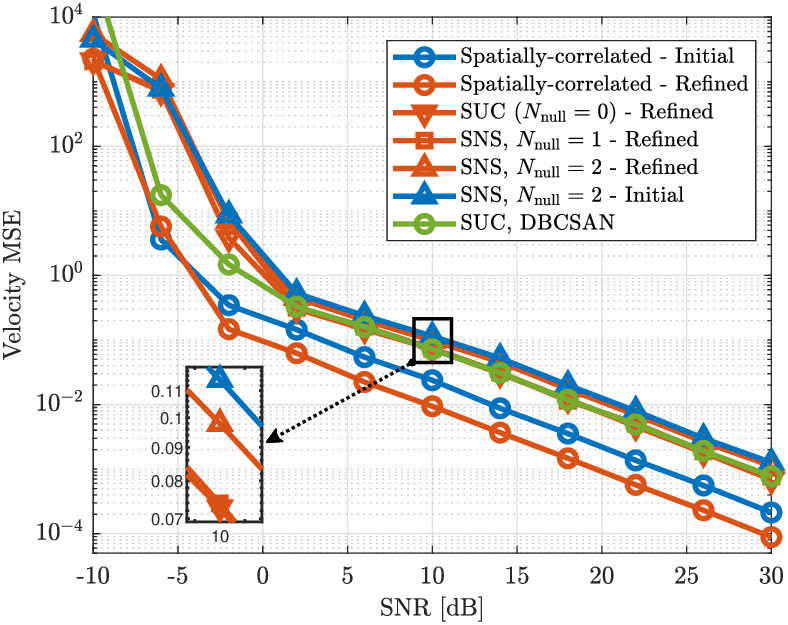}}
\end{center}
\vspace*{-4mm}
\caption{The MSE performance of VIBS regarding (a) position, and (b) velocity.}
\label{fig:MSE_dv} 
\end{figure}

Based on the results of parameter estimation, the performance of the proposed VIBS paradigm is investigated under various channel environments.
Fig.~\ref{fig:MSE_dv} demonstrates the mean square error (MSE) of both positioning and velocity measurement, which are respectively defined as ${\mathbb E}\left\{\sum\nolimits_{l = 1}^L {\left\| {{{\widehat {\bf{p}}}_l} - {{\bf{p}}_l}} \right\|} _2^2\right\}$ and ${\mathbb E}\left\{\sum\nolimits_{l = 1}^L {\left\| {{{\widehat {\bf{v}}}_l} - {{\bf{v}}_l}} \right\|} _2^2\right\}$ with ${{\widehat {\bf{p}}}_l}$ (${{\widehat {\bf{v}}}_l}$) being the estimate of ${{{\bf{p}}}_l}$ (${{{\bf{v}}}_l}$). It can be observed from Fig.~\ref{fig:MSE_dv} that the proposed positioning scheme can achieve sub-meter level accuracy (i.e., MSE of positioning is less than $0$\,dB) within the practical SNR range, and the GDA in Algorithm~\ref{ALG2} can improve the performance of positioning and velocity measurement. This demonstrates the effectiveness of the proposed VIBS paradigm.
The robustness of the proposed scheme against the hostile channel environments is also evident in Fig.~\ref{fig:MSE_dv}. The performance loss under practical SUC channels, compared to the unrealistic spatially-correlated channels, is acceptable. Moreover, the performance loss becomes negligible as the number of null paths (i.e., $N_{\rm null}$) increases from $0$ to $2$, which indicates that the proposed scheme is almost immune to SNS.
As the benchmark, we introduce the near-field sensing model in \cite{wang2024performance} and solve it via the polar-domain algorithm \cite{cui2022channel}, which is marked by ``NISE'' in Fig.~\ref{fig:MSE_dv}\,(a). Note that this benchmark heavily relies on the spatially-correlated channels, rendering it ineffective under the SUC channels and SNS, as shown in Fig.~\ref{fig:MSE_dv}\,(a). Also note that this benchmark does not consider the velocity measurement, while the proposed method can effectively estimate the three-dimensional velocity (rather than the radial velocity only \cite{gao2022integrated}), as shown in Fig.~\ref{fig:MSE_dv}\,(b). 
{\color{red}Fig.~\ref{fig:MSE_dv} also compares the DBSCAN clustering algorithm with the K-means method. Observe that two algorithms perform almost identically, which indicates both centroid-based (K-means) and density-based (DBSCAN) algorithms can accurately classify the estimates even in the presence of noise.}

\begin{figure}[!t]
\captionsetup{font={footnotesize,color={red}}, name = {Fig.}, labelsep = period}
\centering
\includegraphics[width=0.49\textwidth]{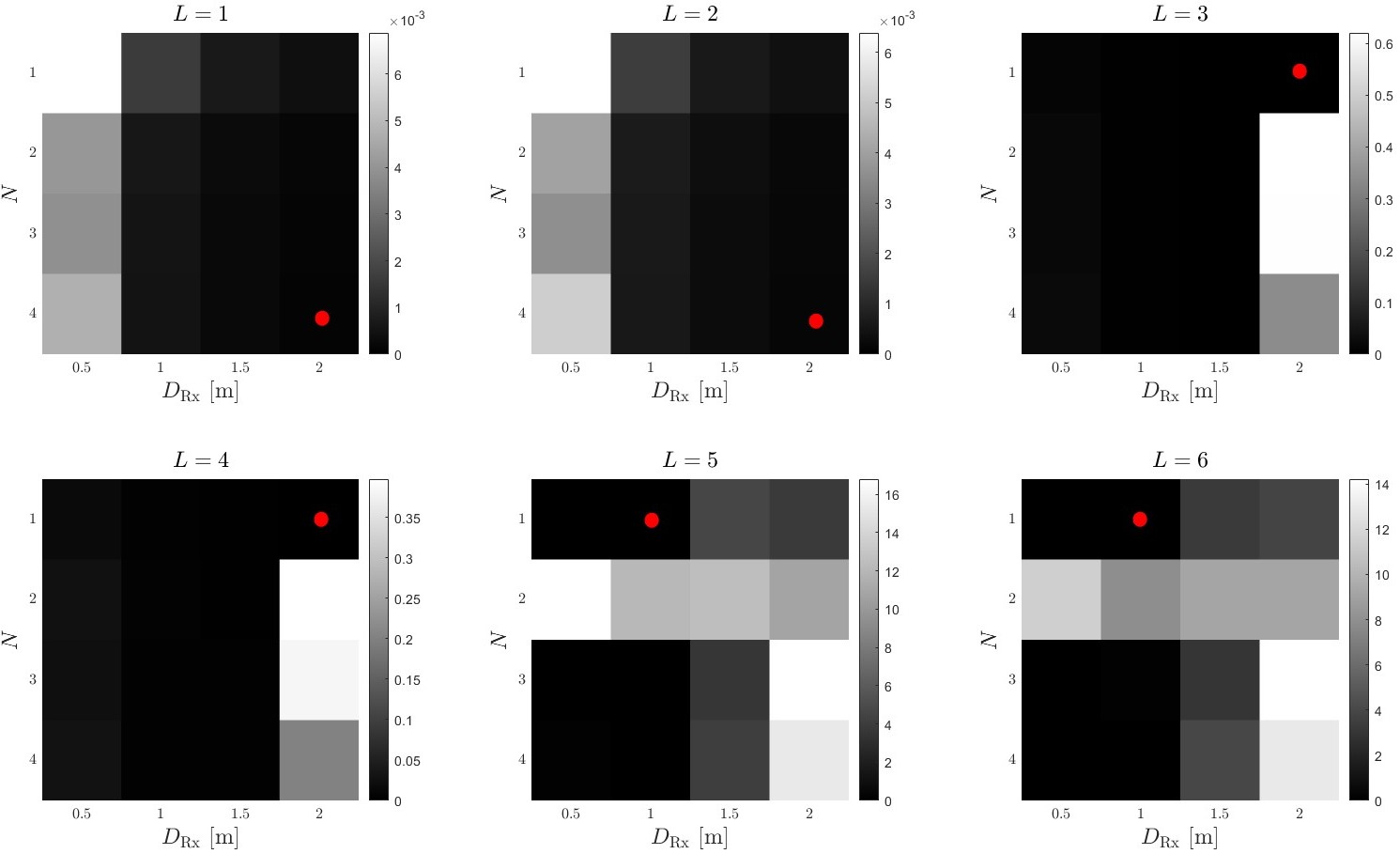}
\caption{Phase transition in positioning MSE versus the number of DSAs/DSSs $N$ and the sensing receiver aperture $D_{\rm Rx}$, under different numbers of targets $L$. The darker cell corresponds to the lower MSE. The best combination of $N$ and $D_{\rm Rx}$ is marked by the red dot for each $L$. SNR = $10$\,dB.}
\label{fig:PT} 
\end{figure}

{\color{red}
We further evaluate the impact of transceiver configuration on sensing performance in Fig.~\ref{fig:PT}, where the phase transition in MSE performance with respect to the number of DSAs/DSSs $N$ and the sensing receiver aperture $D_{\rm Rx}$ is presented.
For simplicity, we focus on the positioning MSE only, and the velocity measurement performance follows the same trend.
We can observe that the impact of transceiver configuration on sensing tasks is quite complicated as $L$ varies, which can be explained from the following perspectives. Firstly, when $L$ is small (e.g., $L = 1$ or $2$), increasing either $N$ or $D_{\rm Rx}$ enhances the spatial diversity of the sensing system by providing more diversified observations of each target, thereby improving the sensing performance. Secondly, increasing $N$ degrades the transmit power allocated to each DSS. This degradation will dominate the sensing performance when $L > 2$, making further increases in $N$ ineffective for improving sensing performance. Thirdly, when $L$ is relatively large (e.g., $L = 5$ or $6$), increasing either $N$ or $D_{\rm Rx}$ will make the parameter estimates ``crowded'' in the range-Doppler plane (see Fig.~\ref{fig:PrePro}) and thus it imposes challenges in classifying the estimates via the clustering algorithm. As a result, the sensing performance degrades sharply with $N$ and/or $D_{\rm Rx}$ in such scenarios.
Fig.~\ref{fig:PT} provides insight in choosing appropriate transceiver configuration for different sensing scenarios. In particular, for scenarios where the distribution of targets exhibits significant sparsity, $N$ or/and $D_{\rm Rx}$ can be increased for better sensing performance, while for the scenarios with more targets, $N$ or/and $D_{\rm Rx}$ should be moderately reduced to avoid negative effects.
}

\subsubsection{Communication performance}

\begin{figure}[tp!]
\vspace*{-4mm}
\captionsetup{font={footnotesize,color={black}}, name = {Fig.}, singlelinecheck=off, labelsep = period}
\begin{center}
\subfigure[]{\includegraphics[width=0.24\textwidth]{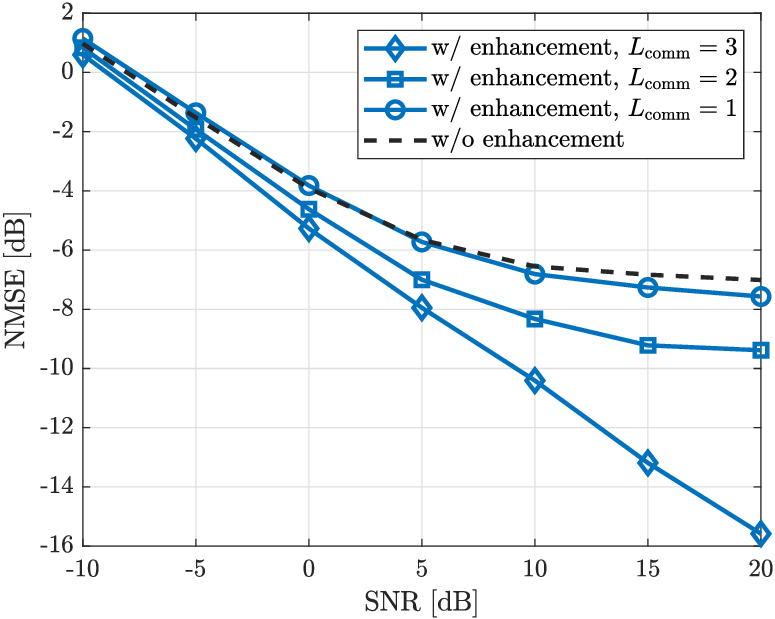}}
\subfigure[]{\includegraphics[width=0.24\textwidth]{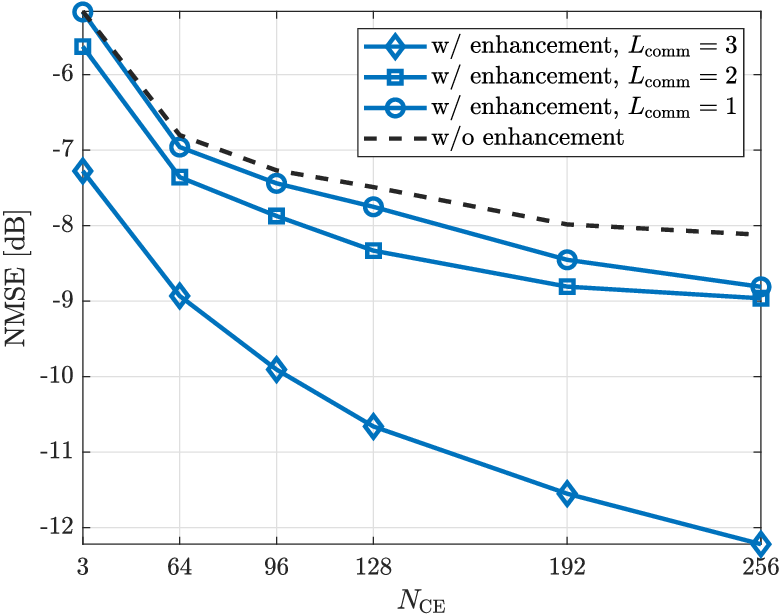}}
\end{center}
\vspace*{-4mm}
\caption{The NMSE performance of CE and its enhancement varying with (a) SNR given $N_{\rm CE} = 128$, and (b) $N_{\rm CE}$ given SNR $= 10$\,dB.}
\label{fig:CE} 
\vspace*{-1mm}
\end{figure}

To evaluate the proposed communication CE scheme, the normalized MSE (NMSE) between the true and estimated communication channels, defined as
\begin{align} 
  {\text{NMSE}} = \mathbb{E}\left\{ {\frac{{\sum\nolimits_{g = 0}^{G - 1} {\left\| {{\bf{\widehat h}}_g^{{\rm{Fd}}} - {\bf{h}}_g^{{\rm{Fd}}}} \right\|^2} }}{{\sum\nolimits_{g = 0}^{G - 1} {\left\| {{\bf{h}}_g^{{\rm{Fd}}}} \right\|^2} }}} \right\},
\end{align}
is adopted as the performance metric,
where ${\bf{\widehat h}}_g^{{\rm{Fd}}}$ denotes the estimated communication channel obtained. We introduce $L_{\rm comm}$ to represent the number of common scatterers shared between the sensing and communication channels. In other words, $L_{\rm comm}$ scatterers from the sensing channel are reused in the communication channel, while the remaining scatterers are distinct. The sensing results are obtained from the proposed VIBS method.
As illustrated in Fig.~\ref{fig:CE}, the proposed frequency-domain near-field CE scheme for UM-MIMO systems performs effectively under the OCDM waveform, achieving an NMSE below $-8$\,dB even when the pilot overhead ($256$) is only half the number of antennas ($512$).
More importantly, a significant CE performance improvement is observed when incorporating the proposed sensing-assisted process. The enhancement becomes more pronounced as the number of common scatterers $L_{\rm comm}$ increases, indicating stronger communication–sensing correlation leads to more accurate CE.

\begin{figure}[!t]
\captionsetup{font={footnotesize}, name = {Fig.}, labelsep = period}
\centering
\includegraphics[width=0.35\textwidth]{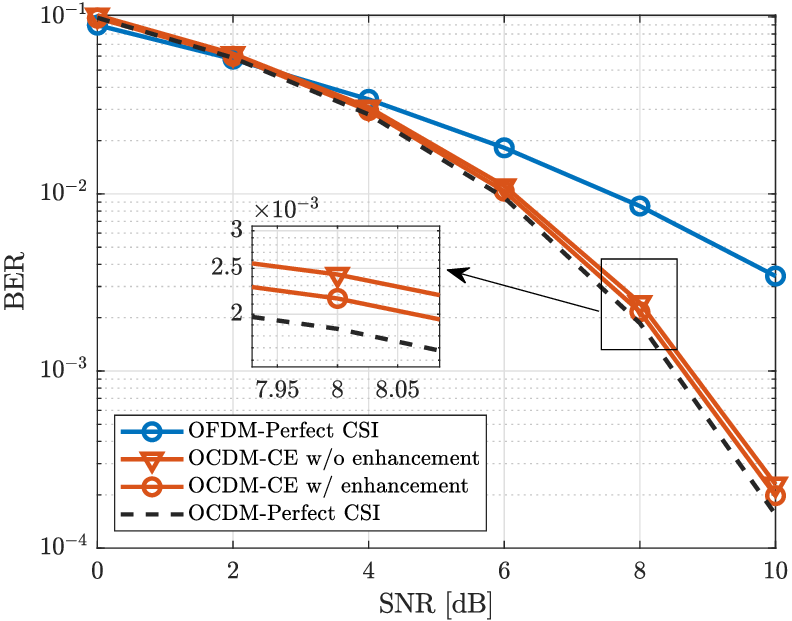}
\caption{Downlink communication BER performance evaluation for the proposed architecture.}
\label{fig:BER} 
\end{figure}

To further verify the communication performance in UM-MIMO and OCDM system, Fig.~\ref{fig:BER} depicts the bit error rate (BER) performance under $N_{\rm CE} = 64$. We adopt the $16$-QAM modulation, and the MMSE equalizers. For the UM-MIMO beamforming design, the channel inversion method is adopted at the ISAC station based on the perfect or estimated channel vectors.
The traditional OFDM is included as the comparison.
As observed from Fig.~\ref{fig:BER}, OCDM has much better BER performance than OFDM. This is because the diversity of OFDM is limited by $1$ in the frequency domain due to its frequency-domain orthogonality, resulting in poor BER performance. OCDM, however, has a better BER performance benefiting from the chirp spectrum spread in the frequency domain, while its transceiver complexity is only little higher \cite{ouyang2016orthogonal}.
Also note that the proposed sensing-assisted CE scheme bring about BER improvement for the UM-MIMO and OCDM, which validates the effectiveness of the proposed near-field ISAC scheme.

%% file: Sec/Sec7_Conclusion.tex
\section{Conclusion}\label{S7}

A novel near-field ISAC solution that integrates UM-MIMO and OCDM techniques has been presented. The proposed ISAC architecture utilizes an OCDM transmitter with UM-MIMO and a co-located FMCW-based sensing receiver, offering efficient dual-functionality. By transmitting DSSs each through a DSA within the UM-MIMO, the system can estimate the ranges and velocities of the near-field targets associated with each transmit-receive antenna pairs with significantly reduced hardware complexity. Furthermore, the virtual bistatic sensing paradigm has been proposed, enabling the robust target positioning and achieving three-dimensional velocity measurement. Moreover, the sensing-enhanced CE has been designed, integrating the sensing results into CE to benefit the communication task. Simulation results have verified the effectiveness and superiority of the proposed ISAC scheme.

%% file: Sec/Appendix.tex
\begin{appendices}
{\color{red}
\section{Proof of {\bf Theorem 1}}\label{app:a}

We note that the vacancy occurs at the beginning of the received signals due to the propagation delay, as illustrated in Fig.~\ref{fig:AppA}. Therefore, we assume that the mixers start operating $T_{\rm GI}$ later after the waveforms are transmitted for each sensing symbol, and our focus is solely on the time interval $\left[ \tau,T\right)$.
By substituting the OCDM waveform expression \eqref{equ:sc_permuted} into $s^{\rm IF}_{n,n'} \left( t ; \tau \right)$ defined in \eqref{equ:DefinesIF}, we obtain
\begin{align}
\label{equ:A1}
    s^{\rm IF}_{n,n'} \left( t ; \tau \right)  = {e^{\textsf{j}{\varphi _{n,n'}}}} e^{{\textsf{j}2\pi \left( {\int_0^{t - \tau } {{{\widetilde \phi }_{k_{n'}}}\left( x \right){\rm{d}}x}  - \int_0^t {{{\widetilde \phi }_{k_n}}\left( x \right){\rm{d}}x} } \right)}},
\end{align}
where 
\begin{align}
\label{equ:A2}
    {\varphi _{n,n'}} = \frac{\pi }{K}\left( {\left\langle {k_n - K/2} \right\rangle _K^2 - \left\langle {k_{n'} - K/2} \right\rangle _K^2} \right).
\end{align}
To cope with the modulus operation in ${\widetilde \phi}_{k_n}\left( x \right)$ and ${\widetilde \phi}_{k_{n'}}\left( x \right)$, we need to divide the time interval of interest ${\left[ {\tau,T} \right)}$ into several pieces and discuss them separately. Given that the discontinuity of $\widetilde \phi _k^{}\left( x \right)$ occurs at $x = k/B$, we should focus on two special points $t = k_n/B$ and $t = k_{n'}/B+\tau$ in \eqref{equ:A1}.
Considering first the case where $\tau < \min \left( {k_n/B,k_{n'}/B + \tau } \right)$ and $T > \max \left( {k_n/B,k_{n'}/B + \tau } \right)$, the following discussions are in order.

\begin{figure}[tp!]
\captionsetup{font={footnotesize,color={red}}, name = {Fig.}, singlelinecheck=off, labelsep = period}
\begin{center}
\subfigure[]{\includegraphics[width=0.23\textwidth]{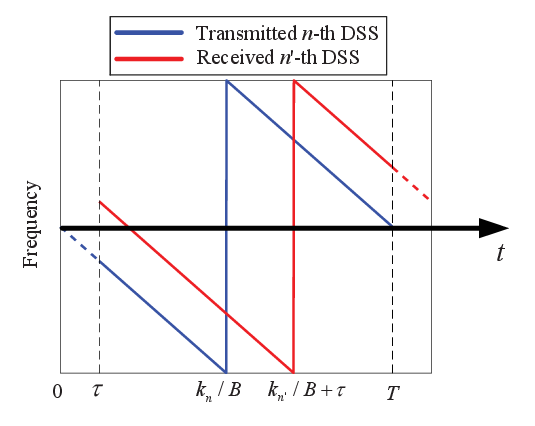}}
\subfigure[]{\includegraphics[width=0.23\textwidth]{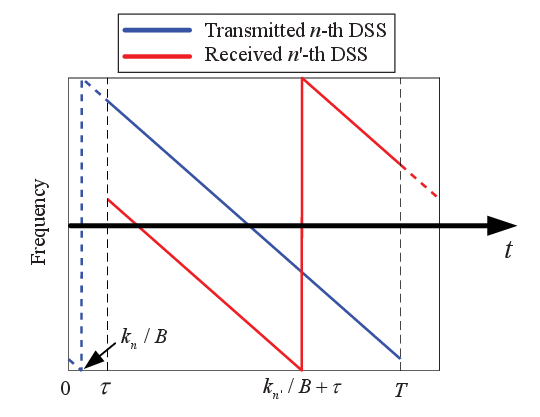}}
\subfigure[]{\includegraphics[width=0.23\textwidth]{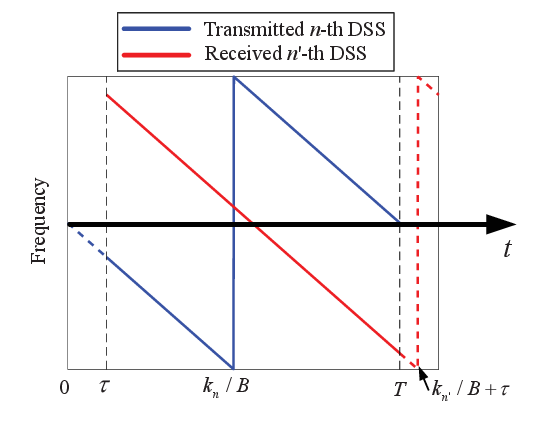}}
\subfigure[]{\includegraphics[width=0.23\textwidth]{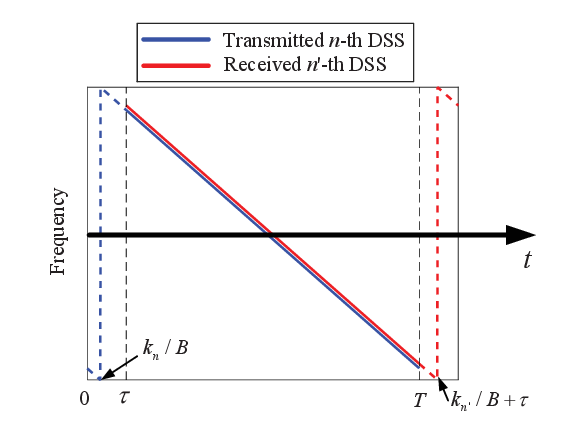}}
\end{center}
\caption{The instantaneous frequency of the transmitted and received DSSs.
(a) The case where ${{\cal T}_{n,n'}^{\rm{I}}\left( \tau  \right)} \neq \emptyset$, ${{\cal T}_{n,n'}^{\rm{II}}\left( \tau  \right)} \neq \emptyset$, and ${{\cal T}_{n,n'}^{\rm{III}}\left( \tau  \right)} \neq \emptyset$.
(b) The case where ${{\cal T}_{n,n'}^{\rm{I}}\left( \tau  \right)} = \emptyset$.
(c) The case where ${{\cal T}_{n,n'}^{\rm{III}}\left( \tau  \right)} = \emptyset$.
(d) The case where ${{\cal T}_{n,n'}^{\rm{I}}\left( \tau  \right)} ={{\cal T}_{n,n'}^{\rm{III}}\left( \tau  \right)} = \emptyset$.
}
\label{fig:AppA}
\end{figure}

\begin{itemize}
    \item $\tau \le t < \min \left( {k_n/B,k_{n'}/B + \tau } \right)$. Accordingly, the phase term in \eqref{equ:A1} can be written as
    \begin{align}
    \label{equ:A3}
        & 2\pi \left( \int_0^{t - \tau } {{{\widetilde \phi }_{{k_{n'}}}}\left( x \right){\rm{d}}x}  - \int_0^t {{{\widetilde \phi }_{{k_n}}}\left( x \right){\rm{d}}x} \right) \nonumber \\
         = & 2\pi \int_0^{t - \tau } {\left( { - {Bx}/{T} + {{{k_{n'}}}}/{T} - {B}/{2}} \right){\rm{d}}x} \nonumber \\
         & \quad \quad \quad - 2\pi \int_0^t {\left( { - {Bx}/{T} + {{{k_n}}}/{T} - {B}/{2}} \right){\rm{d}}x} \nonumber \\
         = & 2\pi \left( {\frac{{B\tau }}{T} + \frac{{{k_{n'}} - {k_n}}}{T}} \right)t + \varphi _n^{\rm{I}}\left( \tau  \right),
    \end{align}
    where
    \begin{align}
        \label{equ:A4}
        \varphi _n^{\rm{I}}\left( \tau  \right) = - \frac{B\pi}{{T}}{\tau ^2} - \left( {\frac{{{2\pi k_n}}}{T} - {B\pi}} \right)\tau.
    \end{align}

    \item $\min \left( {k_n/B,k_{n'}/B + \tau } \right) \le t < \max \left( {k_n/B,k_{n'}/B + \tau } \right)$. If ${k_n}/B < {k_{n'}}/B + \tau$, that is $k_n-k_{n'}<B\tau$, then we have \eqref{equ:A5} shown at the top of the page. Similarly, if $k_n-k_{n'} \ge B\tau$, we can obtain the following results

    \begin{figure*}
        \color{red}
        \begin{align}
        \label{equ:A5}
            2\pi \left( \int_0^{t - \tau } {{{\widetilde \phi }_{{k_{n'}}}}\left( x \right){\rm{d}}x}  - \int_0^t {{{\widetilde \phi }_{{k_n}}}\left( x \right){\rm{d}}x} \right) = & 2\pi \left( \int_0^{t - \tau } {{{\widetilde \phi }_{{k_{n'}}}}\left( x \right){\rm{d}}x}  -  \int_0^{{k_n}/B} {{{\widetilde \phi }_{{k_n}}}\left( x \right){\rm{d}}x}  -  \int_{{k_n}/B}^t {{{\widetilde \phi }_{{k_n}}}\left( x \right){\rm{d}}x} \right) \nonumber \\
             = & 2\pi \left( {\frac{{B\tau }}{T} + \frac{{{k_{n'}} - {k_n}}}{T} - B} \right)t + \varphi _n^{\rm{I}}\left( \tau  \right) + 2\pi{k_n}.
        \end{align}
        \hrule
    \end{figure*}

    \begin{align}
        \label{equ:A6}
        & 2\pi \left(\int_0^{t - \tau } {{{\widetilde \phi }_{{k_{n'}}}}\left( x \right){\rm{d}}x}  - \int_0^t {{{\widetilde \phi }_{{k_n}}}\left( x \right){\rm{d}}x} \right) \nonumber \\
        = & 2\pi \left( {\frac{{B\tau }}{T} + \frac{{{k_{n'}} - {k_n}}}{T} + B} \right)t + \varphi _n^{\rm{I}}\left( \tau  \right) - 2\pi B\tau  - 2\pi {k_{n'}}.
    \end{align}

    Furthermore, the results \eqref{equ:A5} and \eqref{equ:A6} can be re-written in a more compact form as
    \begin{align}
        \label{equ:A7}
        & 2\pi \left( \int_0^{t - \tau } {{{\widetilde \phi }_{{k_{n'}}}}\left( x \right){\rm{d}}x}  - \int_0^t {{{\widetilde \phi }_{{k_n}}}\left( x \right){\rm{d}}x} \right)\nonumber \\
        = & 2\pi \left( {\frac{{B\tau }}{T} + \frac{{{k_{n'}} - {k_n}}}{T} + B\Delta_{n,n'}\left( \tau \right) } \right)t + \varphi _{n,n'}^{\rm{II}}\left( \tau  \right),
    \end{align}
    where
    \begin{align}
        \label{equ:A8}
        \Delta_{n,n'}\left( \tau \right) =  \left\{ {\begin{array}{*{20}{l}}
        { - 1,}&{{k_n} - {k_{n'}} < B\tau }\\
        {1,}&{{k_n} - {k_{n'}} \ge B\tau }
        \end{array}} \right.,
    \end{align}
    and
    \begin{align}
    \label{equ:A9}
        \varphi _{n,n'}^{{\rm{II}}}\left( \tau  \right) = \varphi _n^{\rm{I}}\left( \tau  \right) - \pi \left[ {{1 + {\Delta _{n,n'}}\left( \tau  \right)}} \right] B\tau .
    \end{align}
    Note that the phase terms $2\pi k_n$ in \eqref{equ:A5} and $2\pi k_{n'}$ in \eqref{equ:A6} are neglected, since they will not influence the IF signal.

    \item $\max \left( {{k_n}/B,{k_{n'}}/B + \tau } \right) \le t < T$. We obtain the result in \eqref{equ:A10} shown at the top of the page.

    \begin{figure*}
        \color{red}
        \begin{align}
        \label{equ:A10}
            & 2\pi \left( \int_0^{t - \tau } {{{\widetilde \phi }_{{k_{n'}}}}\left( x \right){\rm{d}}x}  - \int_0^t {{{\widetilde \phi }_{{k_n}}}\left( x \right){\rm{d}}x} \right) \nonumber \\
            = & 2\pi \int_0^{k_{n'}/B } {{{\widetilde \phi }_{{k_{n'}}}}\left( x \right){\rm{d}}x} + 2\pi \int_{k_{n'}/B}^{t - \tau } {{{\widetilde \phi }_{{k_{n'}}}}\left( x \right){\rm{d}}x}  - 2\pi \int_0^{{k_n}/B} {{{\widetilde \phi }_{{k_n}}}\left( x \right){\rm{d}}x}  - 2\pi \int_{{k_n}/B}^t {{{\widetilde \phi }_{{k_n}}}\left( x \right){\rm{d}}x} \nonumber \\
             = & 2\pi \left( {\frac{{B\tau }}{T} + \frac{{{k_{n'}} - {k_n}}}{T}} \right)t + \varphi _n^{\rm{I}}\left( \tau  \right) - 2\pi B\tau  + 2\pi \left( {k_n} - {k_{n'}} \right).
        \end{align}
        \hrule
    \end{figure*}
    
\end{itemize}

By plugging the results \eqref{equ:A3}, \eqref{equ:A7}, and \eqref{equ:A10} (neglecting the term $2\pi \left(k_{n}-k_{n'} \right)$) into \eqref{equ:A1}, we obtain the IF signal expression in \eqref{equ:IF}.
For other cases which do not satisfy $\tau < \min \left( {k_n/B,k_{n'}/B + \tau } \right)$ or $T > \max \left( {k_n/B,k_{n'}/B + \tau } \right)$, we divide the interval of interest $\left[ \tau,T \right)$ into three non-overlapped sets ${{\cal T}_{n,n'}^{\rm{I}}\left( \tau  \right)}$, ${{\cal T}_{n,n'}^{\rm{II}}\left( \tau  \right)}$, and ${{\cal T}_{n,n'}^{\rm{III}}\left( \tau  \right)}$, as defined in \eqref{equ:IF}.
It can be readily verified that when $t$ is drawn from ${{\cal T}_{n,n'}^{\rm{I}}\left( \tau  \right)}$, ${{\cal T}_{n,n'}^{\rm{II}}\left( \tau  \right)}$, and ${{\cal T}_{n,n'}^{\rm{III}}\left( \tau  \right)}$, the phase term of the IF signal $s_{n,n'}^{\rm IF}\left(t;\tau\right)$ will have exactly the same behavior as that in \eqref{equ:A3}, \eqref{equ:A7}, and \eqref{equ:A10}, respectively, therefore the IF signal expression in \eqref{equ:IF} is still valid. However, some of the sets ${{\cal T}_{n,n'}^{\rm{I}}\left( \tau  \right)}$, ${{\cal T}_{n,n'}^{\rm{II}}\left( \tau  \right)}$, and/or ${{\cal T}_{n,n'}^{\rm{III}}\left( \tau  \right)}$ may be empty, as exemplified in Figs.~\ref{fig:AppA}(b)--(d).
This completes the proof of Theorem~\ref{Theorem1}.
}



\section{Proof of Theorem~\ref{Theorem3}}\label{app:c}

We assume that there exists $\overline P^{\left(l\right)} \le P^{\left(l\right)}$ antenna pairs that involve the same DSA, whose coordinate is denoted by ${\bf{p}}_{\rm DSA}^{\left(l\right)}$, and we also assume that the coordinates of the corresponding Rx antennas are ${\bf{p}}^{{\rm{Rx}},\left(l\right)}_p$, $p = 1,2,\ldots,\overline P^{\left(l\right)}$. These assumptions cause no loss of generality.
Based on the measurements from these $\overline P^{\left(l\right)}$ antenna pairs (ignoring the error term), we rearrange \eqref{equ:dis} and square the both sides to obtain
\begin{align}\label{equ:C2} 
  \left({\widehat d^{\left(l\right)}_{p}} - {\left\| {{{ {\bf{p}} }_l} - {\bf{p}}_{\rm DSA}^{\left(l\right)}} \right\|_2} \right)^2 = {\left\| {{{{\bf{p}} }_l} - {\bf{p}} _{p}^{{\rm{Rx}},\left(l\right)}} \right\|^2_2}.
\end{align}
Since all the Tx antennas and Rx antennas are assumed to locate on the xy-plane, the last element in ${{\bf{p}}_{\rm DSA}^{\left( l \right)} - {\bf{p}} _p^{{\rm{Rx}},\left( l \right)}}$ is zero. We denote the first and the second elements of ${{\bf{p}}_{\rm DSA}^{\left( l \right)} - {\bf{p}} _p^{{\rm{Rx}},\left( l \right)}}$ as $\Delta _{p,1}^{\left( l \right)}$ and $\Delta _{p,2}^{\left( l \right)}$, respectively. On this basis, \eqref{equ:C2} can be re-written as
\begin{align}\label{equ:C3} 
  c_p^{\left(l\right)} =  \left({\bf w}_p^{\left(l\right)}\right)^{\rm T} {\bm \theta}^{\left(l\right)},
\end{align}
where
$c_p^{\left(l\right)} = {\left( {\widehat d_p^{\left( l \right)}} \right)^2} + \left\| {{\bf{p}}_{\rm DSA}^{\left( l \right)}} \right\|_2^2 - \left\| {{\bf{p}} _p^{{\rm{Rx}},\left( l \right)}} \right\|_2^2$,
${\bf{w}}_p^{\left( l \right)} = 2{\left[ {\Delta _{p,1}^{\left( l \right)},\Delta _{p,2}^{\left( l \right)},\widehat d_p^{\left( l \right)}} \right]^{\rm{T}}}$, 
and ${\bm \theta}^{\left(l\right)} = \left[ {{x}_{l}},{{y}_{l}},{{\left\| {{{{\bf{p}} }_l} - {\bf{p}}_{\rm DSA}^{\left(l\right)}} \right\|}_2} \right]^{\rm{T}}$.
Stacking \eqref{equ:C3} for $p = 1,2,\ldots,\overline P^{\left(l\right)}$ yields ${\bf c}^{\left(l\right)} = {\bf W}^{\left(l\right)} {\bm \theta}^{\left(l\right)}$,
where ${{\bf{c}}^{\left( l \right)}} = {\left[ {c_1^{\left( l \right)},c_2^{\left( l \right)},\ldots,c_{{{\bar P}^{\left( l \right)}}}^{\left( l \right)}} \right]^{\rm{T}}} \in \mathbb{R}^{\overline P^{\left(l\right)}}$ and ${\bf{W}}_{}^{\left( l \right)} = {\left[ {{\bf{w}}_1^{\left( l \right)},{\bf{w}}_2^{\left( l \right)},\ldots,{\bf{w}}_{{{\bar P}^{\left( l \right)}}}^{\left( l \right)}} \right]^{\rm{T}}} \in \mathbb{R}^{{\overline P^{\left(l\right)}} \times 3}$. Obviously, if $\overline P^{\left(l\right)} \ge 3$, ${\bm \theta}^{\left(l\right)}$ can be uniquely obtained as ${\bm \theta}^{\left(l\right)} = \left({\bf W}^{\left(l\right)}\right)^{\dag}{\bf c}^{\left(l\right)}$,
and thus ${{\bf{p}}_l} = \left[{x}_{l},{y}_{l},{z}_{l} \right]^{\rm T}$ can be accordingly obtained without ambiguity. Specifically, $x_l$ ($y_l$) is the first (second) element of ${\bm \theta}^{\left(l\right)}$, while $z_l$ can be uniquely extracted from the third element of ${\bm \theta}_p^{\left(l\right)}$ provided that $x_l$ and $y_l$ are known.
\end{appendices}